\documentclass[twocolumn,english,aps,prl,superscriptaddress,longbibliography,nofootinbib]{revtex4-2}
\usepackage[T1]{fontenc}
\usepackage[latin9]{inputenc}
\usepackage{color}
\usepackage{babel}
\usepackage{bm}
\usepackage{amsmath}
\usepackage{amssymb}
\usepackage{stmaryrd}
\usepackage{graphicx}
\PassOptionsToPackage{normalem}{ulem}
\usepackage{ulem}
\usepackage[unicode=true,
 bookmarks=false,
 breaklinks=false,pdfborder={0 0 1},backref=false,colorlinks=true]
 {hyperref}
\hypersetup{
 urlcolor=green,linkcolor=blue,citecolor=red}

\makeatletter

\providecommand{\tabularnewline}{\\}

\usepackage{babel}

\makeatother

\begin{document}
\title{Generation of entanglement from mechanical rotation}
\author{Marko Toro\v{s}}
\affiliation{School of Physics and Astronomy, University of Glasgow, Glasgow, G12
8QQ, United Kingdom}
\author{Marion Cromb}
\affiliation{School of Physics and Astronomy, University of Glasgow, Glasgow, G12
8QQ, United Kingdom}
\author{Mauro Paternostro}
\affiliation{Centre for Quantum Materials and Technologies, School
of Mathematics and Physics, Queen's University, Belfast BT7 1NN, United
Kingdom}
\author{Daniele Faccio}
\affiliation{School of Physics and Astronomy, University of Glasgow, Glasgow, G12
8QQ, United Kingdom}
\begin{abstract}
Many phenomena and fundamental predictions, ranging from Hawking radiation
to the early evolution of the Universe rely on the interplay between
quantum mechanics and gravity or more generally, quantum mechanics
in curved spacetimes. However, our understanding is hindered by the
lack of experiments that actually allow us to probe quantum mechanics
in curved spacetime in a repeatable and accessible way. Here we propose
an experimental scheme for a photon that is prepared in a path superposition
state across two rotating Sagnac interferometers that have different
diameters and thus represent a superposition of two different spacetimes.
We predict the generation of genuine entanglement even at low rotation
frequencies and show how these effects could be observed even due
to the Earth's rotation. These predictions provide an accessible platform
in which to study the role of the underlying spacetime in the generation
of entanglement. 
\end{abstract}
\maketitle
\noindent \textbf{\textit{Introduction.--}} Our understanding of
the physical world rests on two theories constructed at the beginning
of the $20^{\textrm{th}}$ century. Quantum mechanics arose out of
the necessity to explain new results coming from deceitfully simple
experiments~\citep{pais1986inward}, whilst general relativity emerged
by recognizing the profound equivalence between inertial and gravitational
effects~\citep{pais1984subtle}. Yet, despite their numerous successes,
we have little experimental evidence about the regime where the two
theories meet. On the one hand, quantum mechanics is well tested in
the domain of elementary particles up to the scale of atoms and macromolecules~\citep{fein2019quantum},
while, on the other hand, the experimental evidence for gravitational
effects is mostly limited to much larger length scales~\citep{will2014confrontation}.

Nonetheless, over the decades a handful of experiments began testing
quantum systems in the underlying spacetime. Among the most notable
are the seminal works on neutron interferometry in the Earth's gravitational
field~\citep{colella1975observation,werner1979effect}. These have
led to a series of experiments which probe interference phenomena
in the regime of Newtonian gravity~\citep{nesvizhevsky2002quantum,fixler2007atom,asenbaum2017phase,overstreet2022observation}
as well as to Sagnac interferometers probing non-inertial rotational
motion~\citep{barrett2014sagnac,bertocchi2006single}.

More recently, the development of photonic technologies have enabled
the exploration of entanglement and multi-mode interference at the
quantum-gravity interface. It has been shown that linear accelerations
do not affect two-photon entanglement~\citep{fink2017experimental},
while low-frequency rotations can modify two-photon Hong-Ou-Mandel
interference~\citep{restuccia2019photon}, and that the anti-bunching
signature of entanglement can be concealed or revealed by low-frequency
rotations~\citep{torovs2020revealing,experiment}. These initial
works suggested general relativistic adaptations~\citep{restuccia2019photon}
to satellite-based missions~\citep{yin2017satellite} with new generalizations
under development~\citep{rivera2021outperforming,brady2021frame,kish2022quantum,barzel2022observer,mieling2022measuring}.

{Ring laser gyroscopes have also achieved exquiste
sensitivities in underground facilities~\cite{di2021sensitivity},
with proposals for testing the Lense-Thirring effect~\cite{di2017ginger}
and for constraining theories of gravity~\cite{capozziello2021constraining},
offering an alernative to orbiting cryogenic gyroscopes~\cite{everitt2011gravity}
and to satellite laser ranging~\cite{ciufolini2004confirmation}.
Moreover, Sagnac-based interferometers have been suggested for
detecting gravitational waves from intermediate-mass black hole mergers~\cite{lacour2019sage},
and are the backbone of fundamental and technological applications~\cite{kim2019pulsed,lee2021sagnac}. }

The above experiments and proposals, striking in their own right,
have in common that the degree of entanglement remains unaltered by
the underlying spacetime. Although theoretical calculations are indicating
that entanglement is not an invariant quantity in a general relativistic
setting~\citep{fuentes2005alice,alsing2012observer} any variations
become vanishingly small at low accelerations or in weak gravity~\citep{fink2017experimental}.
All experimental and theoretical results are thus suggesting, at least
in the regime within reach of typical laboratory experiments, that
entanglement remains unaltered by the underlying spacetime.

As we show here, this is not the case: we provide a protocol for \emph{generating
entanglement} in the regime of low accelerations. We exploit a previously
unexplored coupling that arises in non-inertial rotating reference
frames. We will focus on an implementation with photonic systems, which
offers the prospect of an experimental implementation using the path-polarization
degrees of freedom. In particular, we will show that an initially
separable state becomes \emph{maximally entangled} even at low-frequency
of rotations $\sim1~\text{Hz}$, using fibers of length $\sim30~\text{m}$,
and a platform of radius $\sim0.5~\text{m}$. {The
scheme relies on a single photon source, the original use of a
dual-Sagnac interferometer -- which allows for the interpretation of our results in terms
of spacetime superpositions -- and protocols for witnessing quantum
entanglement.} We discuss the implications for {the
quantum-gravity interface},
and conclude by estimating the experimental requirements to test the
generation of entanglement driven by the Earth's daily rotation.

\begin{figure*}
\includegraphics[width=1\textwidth]{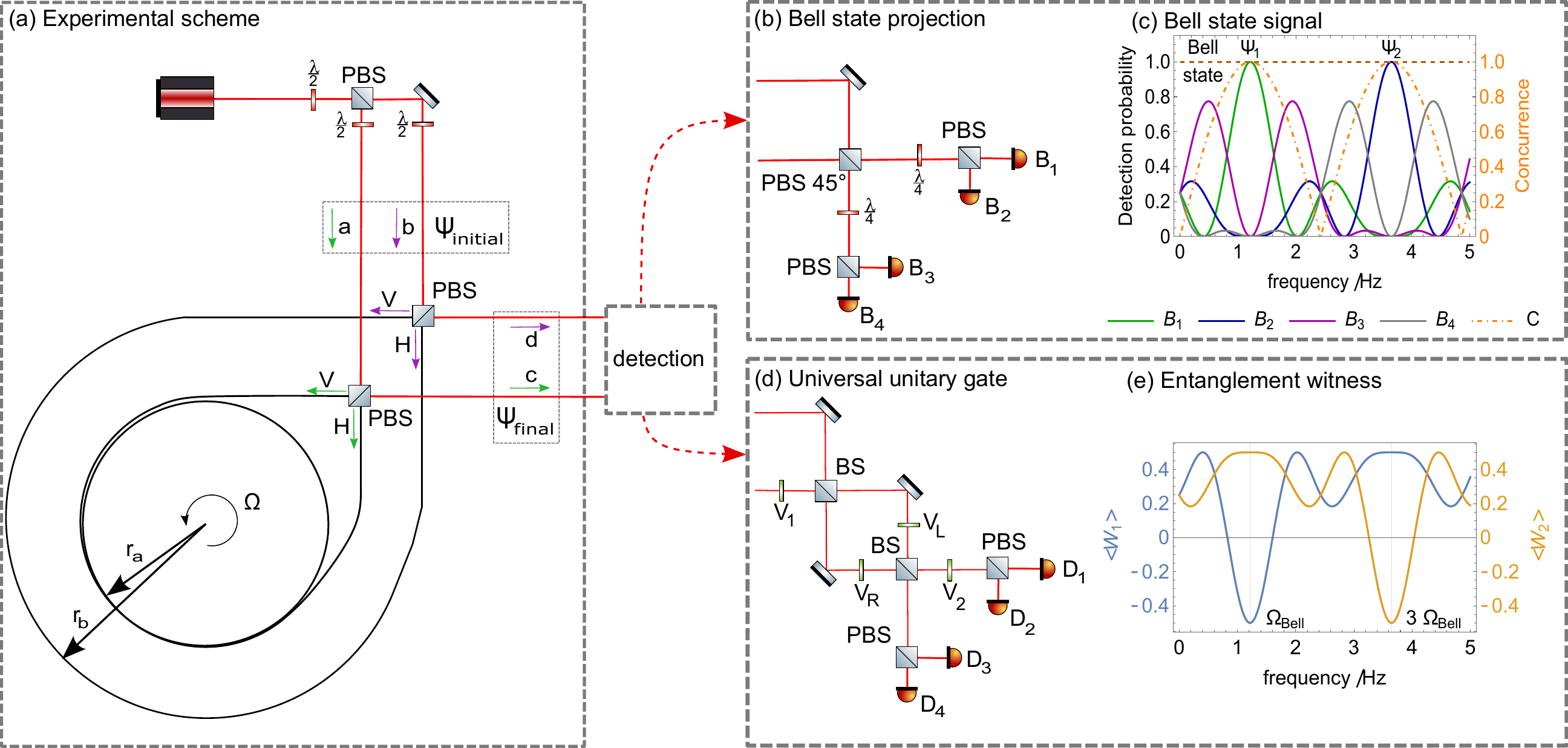}

\caption{Scheme for generating path-polarization entanglement from mechanical
rotation. \textbf{(a) } The scheme consists of a single photon source,
polarizing beam splitters (PBSs), and half-wave plates (HWPs) denoted
by $\lambda/2$.
We consider fiber loops with radii $r_{b}\sim0.5~\text{m}$, $r_{a}\sim r_{b}/2$
(with total length $l\sim2\pi r_{b}N_{b}$ with winding number $N_{b}=10$)
and a photon wavelength $800~\text{nm}$. The experimental setup is
placed on a platform which can be set in rotation with frequency $\Omega$.
The purple and green arrows indicate the paths $a$, $b$, while the
polarization is denoted by $H$, $V$. We start with a separable path-polarization
photon $\vert\psi_{\text{initial}}\rangle$; depending on the frequency
of rotation $\Omega$ the final photon $\vert\psi_{\text{final}}\rangle$
can remain separable or becomes entangled. We can generate a maximally
entangled Bell state by tuning the frequency of the platform to $\Omega_{\text{Bell}}=\pi c^{2}/(2\omega\mathcal{A})$,
where $\omega$ is the mean photon frequency, and $\mathcal{A}=\Delta rl$
is the effective area of the interferometer ($\Delta r=r_{b}-r_{a}$
is the difference of the radii, and $l$ is the path length assumed
to be equal for the $a$, $b$ paths).\textbf{ (b)} The detector $B_{j}$
measures the Bell-state $\psi_{j}$ ($j=1,..,4$). The first PBS is
rotated by $\pi/4$, and the quarter-wave plates (QWPs) are denoted
by $\lambda/4$ (with the fast axis oriented at $\pi/4$ which transforms
circular polarization to linear polarization). \textbf{(c) }The detector
$B1$ ($B2$) measures the maximally entangled Bell-state $\psi_{1}$($\psi_{2}$)
at frequency $\Omega_{\text{Bell}}$($3\Omega_{\text{Bell}}$) where
the concurrence achieves the maximum possible value $C=1$. We find
$\Omega_{\text{Bell}}\sim2\pi\times1.2~\text{Hz}$, which can be readily
achieved with similar photonic setups~\citep{restuccia2019photon}.
\textbf{(d)} Universal unitary gate for single-photon two-qubit states
which can be used for tomographic reconstruction or implementing entanglement
witnesses. The two beam-splitters (BSs) together with the mirrors
form a Mach-Zehnder interferometer, and $V_{j}$($j=1,R,L,2$) denotes
an optical element composed of a HWP, two enclosing QWPs, and a phase
shifter. \textbf{(e)} Optimal entanglement witnesses $\mathcal{W}_{1}$
($\mathcal{W}_{2}$) for the maximally entangled Bell-state $\psi_{1}$($\psi_{2}$)
as a function of rotation frequency. Entanglement is witnessesed when
$\mathcal{W}_{j}<0$ ($j=1,2$). }
\label{fig:experiment} 
\end{figure*}

\noindent \textbf{\textit{Coupling rotations and paths. -- }} The
study of rotating reference frames has led to major breakthroughs
in experimental and theoretical physics, from Sagnac's test of Special
relativity~\citep{sagnac1913ether,sagnac1913preuve}, to Einstein's
General theory of relativity~\citep{einstein1909letter}. In this
work we are interested in dynamical effects that arise from the motion
of quantum systems in a rotating Minkowski spacetime.

The equations of motion can be constructed from two simple observations.
First, non-inertial rotational effects scale with the frequency of
rotation $\bm{\Omega}.$ Second, from the viewpoint of the co-rotating
observer, free-moving objects are rotating around the origin, and
must thus possess an angular momentum $\bm{J}$. We are thus led to
the Hamiltonian term $\sim\bm{\Omega}\cdot\bm{J}$ \citep{korsbakken2004fulling}.
The situation further simplifies for motion on a circle with the axis
of rotation located at the origin of the coordinates~\citep{torovs2020revealing}
\begin{equation}
H_{\text{rot}}=H+\Omega rp,\label{eq:Hamiltonian}
\end{equation}
where $H$ is the usual Hamiltonian that is present already in an
inertial reference frame, and $r$ ($p$) is the radial position (tangential
momentum) of the system.

Let us now consider the motion of photons in such a rotating spacetime.
We can gain an intuitive understanding of Eq.~(\ref{eq:Hamiltonian})
by recalling the relation between energy and momentum~\citep{padgett2008diffraction,barnett2010resolution},
$p=\pm H/(nc)$, where $n$ is the refractive index of the medium,
and $c$ is the speed of light. The Hamiltonian hence transforms to
$H_{\text{rot}}^{(\pm)}=H(1\pm\frac{\Omega r}{nc})$, where $\pm\frac{\Omega r}{nc}$
can be seen as the Doppler shift of the energy due to the rotational
motion (with the sign indicating counter-rotating/co-rotating motion
with respect to the platform). It is precisely this imbalance between
the two directions of motion that is responsible for the Sagnac phase,
$\phi_{s}=(H_{\text{rot}}^{(+)}-H_{\text{rot}}^{(-)})t/\hbar=2H\Omega rt/(\hbar nc)$,
where $t$ is the time of flight. In particular, setting $H=\hbar\omega$
and $t=2\pi r/(c/n)$, we then readily find the usual expression for
the Sagnac phase~\citep{post1967sagnac} 
\begin{equation}
\phi_{s}=\frac{4\Omega\omega\mathcal{A}_{s}}{c^{2}},\label{eq:Sagnac}
\end{equation}
where $\omega$ is the optical frequency, and $\mathcal{A}_{s}=\pi r^{2}$
denotes area enclosed by the interferometer.

The term $\Omega rp$ in Eq.~(\ref{eq:Hamiltonian}) can be also
viewed as a \emph{coupling} between the position, $r$, and the momentum,
$p$, with the values depending on the path followed by the photon.
In a non-rotating reference frame $\Omega=0$ and the coupling vanishes.
As we will see, this coupling can be exploited to entangle the path-polarization
degrees of freedom of photons.

\noindent \textbf{\textit{Experimental scheme. -- }}The experimental
setup is shown in Fig.~\ref{fig:experiment} (a). We consider the
initial state 
\begin{equation}
\vert\psi_{\text{initial}}\rangle=\frac{1}{2}(\vert a\rangle+\vert b\rangle)(\vert H\rangle+\vert V\rangle),\label{eq:initial}
\end{equation}
where $a$, $b$ denote the path, and $H$, $V$ denote the polarization
{(cf. Sec.~A of the Supplementary Material accompanying the manuscript for more details)}.
We note that the initial state is separable into a path part, $(\vert a\rangle+\vert b\rangle)/\sqrt{2}$,
and a polarization part $(\vert H\rangle+\vert V\rangle)/\sqrt{2}$.

We then send the photon into \emph{four} different paths with different
radii and momenta, namely $(r_{a},p)$, $(r_{a},-p)$, $(r_{b},p)$,
and $(r_{b},-p)$. As a consequence, the non-inertial rotational motion,
via the coupling $\Omega rp$ in Eq.~(\ref{eq:Hamiltonian}), induces
four different phases in $\vert\psi_{\text{initial}}\rangle$. In
particular, we find the final state 
\begin{align}
\vert\psi_{\text{final}}\rangle=\frac{1}{2}\Bigl( & \vert a\rangle[e^{-\frac{i\Omega r_{a}pt}{\hbar}}\vert H\rangle+e^{\frac{i\Omega r_{a}pt}{\hbar}}\vert V\rangle]\nonumber \\
 & +\vert b\rangle[e^{-\frac{i\Omega r_{b}pt}{\hbar}}\vert H\rangle+e^{\frac{i\Omega r_{b}pt}{\hbar}}\vert V\rangle]\Bigr),\label{eq:psifinal}
\end{align}
where $t={nl}/{c}$ is the the flight time with $n$ the refractive
index of the fiber, $l$ the length of the fibers (assumed equal for
the two paths), $c$ the speed of light, $p=E/(nc)=\hbar\omega/(nc)$
the momentum, and $\omega$ the photon frequency. We note that the
phases can be rewritten as 
\begin{equation}
\phi_{j}\equiv\frac{\Omega r_{j}pt}{\hbar}=\frac{\Omega\omega\mathcal{A}_{j}}{c^{2}},\;(j=a,b)
\end{equation}
where $\mathcal{A}_{j}=r_{j}l$ is an effective area of the loop.
{The phases $\phi_{j}$ can thus be seen as variants
of the Sagnac phase introduced in Eq.~(\ref{eq:Sagnac}). Importantly,
such phases do not depend on the refractive index of the medium, indicating
that the effect highlighted here does not stem from light-matter coupling in the
fibers but rather that it is rooted in relativity~\cite{post1967sagnac,anderson1994sagnac,malykin2000sagnac}.}

Let us consider first the case $\Omega=0$. We note that the final
state in Eq.~(\ref{eq:psifinal}) reduces to the initial state in
Eq.~(\ref{eq:initial}) and thus a non-rotating platform has no effect
on entanglement, as expected. Conversely when $\Omega\neq0$ the state
in Eq.~(\ref{eq:psifinal}) will in general become entangled (as
we can no longer write it as the product of the path and polarization
states). The effect of mechanical rotation is to rotate the polarization
state depending on the photon path $a,b$ -- as we will see the polarization
states in Eq.~(\ref{eq:psifinal}) associated to paths $a$ and $b$
can become orthogonal (the overall polarization states in the square
brackets), resulting in a maximally entangled path-polarization state.
This shows that non-inertial rotating motion generates entanglement.

\noindent \textbf{\textit{Spacetime superpositions. -- }}The presented
mechanism for the generation of entanglement{, arising
from the ostensibly relativistic Sagnac effect~\cite{post1967sagnac,anderson1994sagnac,malykin2000sagnac},}
poses fundamental questions about role of the underlying spacetime.
Although the final state in Eq.~(\ref{eq:psifinal}) can be derived
within the framework of quantum field theory in curved spacetime,
the two rotating Sagnac interferometers shown in Fig.~\ref{fig:experiment}
are suggestive of an interpretation in terms of spacetime superpositions.
This can be put in a mathematical form by modelling the spacetimes
inside the two fiber loops. The metric is given by~\citep{torovs2020revealing}:

\begin{equation}
ds_{j}^{2}=c^{2}(1-\frac{\Omega^{2}r_{j}^{2}}{c^{2}})dt^{2}-2\Omega r_{j}^{2}dtd\phi-r_{j}^{2}d\phi^{2},\;(j=a,b)\label{eq:metric}
\end{equation}
where the time $t$ and polar angle $\phi$ are the two coordinates,
and the subscripts $a$ ($b$) denote the small (big) fiber loop.
{The 1+1 metric in Eq.~\eqref{eq:metric} is fully
specified by two quantities: the angular frequency of rotation $\Omega$
and the radius $r_{j}$. Together, they encode the specific relative angular momentum $\Omega r_{j}^{2}$ in the $dt\,d\phi$ term,
 and the factor $\Omega^{2}r_{j}^{2}/c^{2}$, responsible
for time dilation, in
the $dt^{2}$ term~\cite{thorne2000gravitation,vallado2001fundamentals}}.
{To account for these two
effects, which are properties of the spacetime and not of the photons,
we introduce the states $\vert r_{j},\Omega\rangle$ ($j=a,b$). }

As the photon enters the two fiber loops we are thus led to consider
the following joint photon-spacetime state:

\noindent 
\begin{equation}
\vert\psi\rangle=\frac{1}{2}(\vert a\rangle\vert r_{a},\Omega\rangle+\vert b\rangle\vert r_{b},\Omega\rangle)(\vert p\rangle\vert H\rangle+\vert-p\rangle\vert V\rangle),\label{eq:initialafter}
\end{equation}
where $\vert r_{j},\Omega\rangle$ ($j=a,b$) denotes the state associated
to the metric in Eq.~\eqref{eq:metric}, and $\vert\pm p\rangle$
is the photon momentum state. The state $\vert a\rangle\vert H\rangle$
appearing in Eqs.~(\ref{eq:initial}) and (\ref{eq:psifinal}) can
be thus seen as a shorthand notation for the photon-spacetime state
$\vert a\rangle\vert r_{a},\Omega\rangle\vert p\rangle\vert H\rangle$
(and similarly for the other three states $\vert a\rangle\vert V\rangle$,
$\vert b\rangle\vert H\rangle$, and $\vert b\rangle\vert V\rangle$).

In particular, the term $\vert a\rangle\vert r_{a},\Omega\rangle+\vert b\rangle\vert r_{b},\Omega\rangle$
in Eq.~\eqref{eq:initialafter} can be interpreted as a superposition
of spacetimes associated to the two Sagnac interferometers. The state
of the metric $\vert r_{j},\Omega\rangle$ ($j=a,b$) will induce
phases proportional to $r_{j}\Omega$, which is a key ingredient in
entangling the path-polarization degrees of freedom of the photon.
Hence, by promoting the metric in Eq.~\eqref{eq:metric} to a quantum
state we have shown that the generation of entanglement can be linked
to the concept reminiscent of quantum reference frames~\citep{aharonov1984quantum,giacomini2019relativistic}.

\noindent \textbf{\textit{Maximizing entanglement. -- }}The state
in Eq.~(\ref{eq:psifinal}) will become maximally entangled when
the overall polarization states in the square brackets of Eq.~(\ref{eq:psifinal})
become orthogonal. This is achieved when the overlap $S$ between
such states reduces to zero. Using the orthonormality of the $\vert H\rangle$,
$\vert V\rangle$ polarization states, we find 

\noindent 
\begin{equation}
\begin{aligned}S & =\frac{1}{2}[e^{i\phi_{a}}\langle H\vert+e^{-i\phi_{a}}\langle V\vert][e^{-i\phi_{b}}\vert H\rangle+e^{i\phi_{b}}\vert V\rangle]\\
 & =\cos(\Omega\omega\Delta rl/c^{2}),
\end{aligned}
\label{eqS}
\end{equation}
where we have set $pt/\hbar=\omega l/c^{2}$ ($\Delta r=r_{b}-r_{a}$
is the difference of the radii, and $l$ is the path length assumed
to be equal for the $a,b$ paths). We thus have $S=0$ (and the path-polarization
state becomes maximally entangled) when the rotation frequency $\Omega$
takes the values 
\begin{equation}
\Omega_{\text{Bell}}\equiv\frac{(2k+1)\pi c^{2}}{2\omega\mathcal{A}},~~~~(k\in\mathbb{Z})\label{eq:bell}
\end{equation}
where $\mathcal{A}=\Delta rl$ is the effective area of the interferometer.
If we set the rotation frequency $\Omega$ to any odd frequency multiple
of $\Omega_{\text{Bell}}$ we will also generate a maximally entangled
Bell state, while if we tune $\Omega$ to any even frequency multiple
of $\Omega_{\text{Bell}}$, the state remains separable. We find the
Bell states 
\begin{equation}
\begin{aligned}\vert\psi_{1}\rangle=\frac{1}{2}\Bigl( & \vert a\rangle[e^{-i\frac{\pi}{2}\frac{r_{a}}{\Delta r}}\vert H\rangle+e^{i\frac{\pi}{2}\frac{r_{a}}{\Delta r}}\vert V\rangle]\\
 & +\vert b\rangle[e^{-i\frac{\pi}{2}\frac{r_{b}}{\Delta r}}\vert H\rangle+e^{i\frac{\pi}{2}\frac{r_{b}}{\Delta r}}\vert V\rangle]\Bigr)
\end{aligned}
\label{eq:psifinal22}
\end{equation}
at $\Omega/\Omega_{\text{Bell}}=...,-7,-3,1,5,...$., and 
\begin{equation}
\begin{aligned}\vert\psi_{2}\rangle=\frac{1}{2}\Bigl( & \vert a\rangle[e^{-i\frac{3\pi}{2}\frac{r_{a}}{\Delta r}}\vert H\rangle+e^{i\frac{3\pi}{2}\frac{r_{a}}{\Delta r}}\vert V\rangle]\\
 & +\vert b\rangle[e^{-i\frac{3\pi}{2}\frac{r_{b}}{\Delta r}}\vert H\rangle+e^{i\frac{3\pi}{2}\frac{r_{b}}{\Delta r}}\vert V\rangle]\Bigr)
\end{aligned}
\label{eq:psifinal33}
\end{equation}
at $\Omega/\Omega_{\text{Bell}}=...,-5,-1,3,7...$. 
Non-inertial rotational motion is thus able to generate two distinct
Bell states $\vert\psi_{1}\rangle$,$\vert\psi_{2}\rangle$ -- from
an initially separable state -- solely by tuning the frequency of
rotation.

\noindent \textbf{\textit{Verfying entanglement. -- }}We can verify
the generation of maximal entanglement using the Bell state projection
scheme~\citep{kim2003single} shown in Fig.~\ref{fig:experiment}(b)
with the experimental signature in Fig.~\ref{fig:experiment}(c).
In particular, we compute the concurrence $C=2\vert\mathfrak{a}_{1}\mathfrak{a}_{4}-\mathfrak{a}_{2}\mathfrak{a}_{3}\vert$
as witness of the degree of entanglement for different rotation frequencies
\citep{hill1997entanglement}, where $\mathfrak{a}_{1}$,..,$\mathfrak{a}_{4}$
denote the four phase factors (including the numerical prefactor $1/2$)
in the order appearing in Eq.~\eqref{eq:psifinal}. When the probability
of detection in $B_{1}$ or $B_{2}$ is unity then the concurrence
reaches the value $C=1$, indicating maximum entanglement (see Supplementary
material B and C for more details).

We can also use the universal unitary gate for single-photon two-qubit
states~\citep{englert2001universal} shown in Fig.~\ref{fig:experiment}(d)
to perform a full tomographic reconstruction or to implement a fidelity-based
optimal entanglement witness~\citep{guhne2009entanglement}. The
entanglement established in the system is validated whenever the entanglement
witness $\mathcal{W}_{1}$ or $\mathcal{W}_{2}$ shown in Fig.~1(e)
acquires a negative value (see Supplementary Material D and E for
more details). {In Sec. F of the Supplementary Material, we
also propose an alternative experimental scheme, based on a single-loop
configuration, that is capable of achieving similar results to those
reported here, but is not susceptible to alignment imperfections between
the two loops (see Sec. G of Supplementary Material for a robustness analysis).}

\noindent \textbf{\textit{Discussion. -- }}We have shown that path-polarization
entanglement can no longer be viewed as an invariant quantity, but
rather requires us to view it as a dynamical quantity that can be
completely altered already by low-frequency rotations.



The literature sometimes makes a distinction between path-polarization
entanglement and multi-photon entanglement. However, one can transfer
intra-photon entanglement of the two separable photons to a two-photon
entangled state using known entanglement swapping protocols resulting
in a polarization-entangled photon pair~\citep{adhikari2010swapping,kumari2019swapping}.
The scheme of Fig.~\ref{fig:experiment} could be thus combined with
entanglement swapping protocols to generate multi-photon entanglement{,
offering an intriguing alternative to electrodynamic protocols~\cite{chen2016photon,chen2017dissipation,chen2021two}.}

The developed scheme is also not a peculiarity of photonic systems
or rotations, but readily offers the possibility of adaptations and
generalizations. It can be adapted to matter-wave interferometers,
as the Hamiltonian in Eq.~(\ref{eq:Hamiltonian}) applies to any
system, whether massless or massive. The scheme could also be modified
to probe other gravitational couplings, such as those involving linear
accelerations and spacetime curvature, although such effects are typically
weaker~\citep{zych2012general} and would require a dedicated space
mission~\citep{yin2017satellite}. Here we have focused on the strongest
effect that emerges directly in a rotating reference frame in a Minkowski
spacetime. The proposed scheme is thus also fundamentally different
from gravitationally induced entanglement between two massive systems~\citep{bose2017spin,marletto2017gravitationally},
which{{} }{{}
will test the quantum nature of} perturbations around the Minkowski
spacetime~\cite{marshman2020locality,danielson2022gravitationally,bose2022mechanism,christodoulou2022locally}.

{Other tests of the quantum-gravity interface aim to
probe specific quantum gravity phenomena~\cite{amelino2013quantum}
such as holographic fluctuations in spacetime with Michelson interferometers~\cite{chou2017interferometric,vermeulen2021experiment,verlinde2021observational},
modified commutation relations with optomechanical setups~\cite{pikovski2012probing},
and energy dispersion in astronomical observations~\cite{amelino1998tests}.
There are also a number of experiments which are testing for possible
violations of Lorentz invariance~\cite{kostelecky2011data} which
could arise in certain models of quantum gravity~\cite{mattingly2005modern},
and there is ongoing effert to test the Penrose wavefunction collapse~\cite{penrose1996gravity,donadi2021underground,PhysRevLett.129.080401}
as well as other non-standard mechanisms of gravitational decoherence~\cite{bassi2017gravitational}. }

In summary, the scheme presented in this work addressed a hitherto
unexplored process for the dynamical generation of entanglement from
the underlying spacetime, which lends itself to a suggestive interpretation
in terms of spacetime superpositions.{{} Its core strength
is that the predictions do not depend on specific models of quantum
gravity but only on elementary notions of quantum field theory in
curved spacetime.}  {Furthermore,
it} is based on well-established tools from quantum optics, and it
can be readily experimentally implemented using rotational frequency
$\sim1$ Hz, fibers of length $\sim30$ m, and a platform of radius
$\sim0.5$ m, similar to the numbers achieved in Ref.~\citep{restuccia2019photon}.

Even more intriguing is the fact that any rotation, even the Earth's
daily one, may be used to continuously generate entanglement. Setting
$\Omega_{\text{Bell}}\sim\Omega_{\text{Earth}}\sim7\times10^{-5}\text{Hz}$
in Eq.~(\ref{eq:bell}) we find that the area required to generate
maximal entanglement is about $\sim0.65\text{km}^{2}$, which is comparable
to the interferometer built by Michelson in 1925~\citep{michelson1925effect1,michelson1925effect2}.
It thus appears that testing the generation of entanglement sourced
by the Earth's daily rotation is well within the domain of current
experimental capabilities.

\noindent \textbf{\textit{Acknowledgements. -- }} The authors acknowledge
financial support from the Leverhulme Trust (grants RPG-2020-197 and
RGP-2018-266), the European Union's Horizon 2020 FET-Open project
TEQ (766900), the Royal Society Wolfson Fellowship (RSWF/R3/183013),
the UK EPSRC (grants EP/T028424/1, EP/T00097X/1, EP/W007444/1, EP/R030413/1,
EP/M01326X/1, EP/R030081/1), the Department for the Economy Northern
Ireland under the US-Ireland R\&D Partnership Programme, and the Royal
Academy of Engineering Chair in Emerging Technologies programme.

 \bibliographystyle{unsrt}
\bibliography{bibliography}

\begin{thebibliography}{10}

\bibitem{pais1986inward}
Abraham Pais.
\newblock Inward bound: of matter and forces in the physical world.
\newblock 1986.

\bibitem{pais1984subtle}
Abraham Pais and Stanley Goldberg.
\newblock ``{S}ubtle is the {L}ord...'': {T}he {S}cience and the {L}ife of
  {A}lbert {E}instein, 1984.

\bibitem{fein2019quantum}
Yaakov~Y Fein, Philipp Geyer, Patrick Zwick, Filip Kia{\l}ka, Sebastian
  Pedalino, Marcel Mayor, Stefan Gerlich, and Markus Arndt.
\newblock Quantum superposition of molecules beyond 25 kda.
\newblock {\em Nature Physics}, 15:1242, 2019.

\bibitem{will2014confrontation}
Clifford~M Will.
\newblock The confrontation between general relativity and experiment.
\newblock {\em Living reviews in relativity}, 17(1):1--117, 2014.

\bibitem{colella1975observation}
Roberto Colella, Albert~W Overhauser, and Samuel~A Werner.
\newblock Observation of gravitationally induced quantum interference.
\newblock {\em Physical Review Letters}, 34(23):1472, 1975.

\bibitem{werner1979effect}
SA~Werner, J-L Staudenmann, and R~Colella.
\newblock Effect of earth's rotation on the quantum mechanical phase of the
  neutron.
\newblock {\em Physical Review Letters}, 42(17):1103, 1979.

\bibitem{nesvizhevsky2002quantum}
Valery~V Nesvizhevsky, Hans~G B{\"o}rner, Alexander~K Petukhov, Hartmut Abele,
  Stefan Bae{\ss}ler, Frank~J Rue{\ss}, Thilo St{\"o}ferle, Alexander Westphal,
  Alexei~M Gagarski, Guennady~A Petrov, et~al.
\newblock Quantum states of neutrons in the earth's gravitational field.
\newblock {\em Nature}, 415(6869):297--299, 2002.

\bibitem{fixler2007atom}
Jeffrey~B Fixler, GT~Foster, JM~McGuirk, and MA~Kasevich.
\newblock Atom interferometer measurement of the newtonian constant of gravity.
\newblock {\em Science}, 315(5808):74--77, 2007.

\bibitem{asenbaum2017phase}
Peter Asenbaum, Chris Overstreet, Tim Kovachy, Daniel~D Brown, Jason~M Hogan,
  and Mark~A Kasevich.
\newblock Phase shift in an atom interferometer due to spacetime curvature
  across its wave function.
\newblock {\em Physical Review Letters}, 118(18):183602, 2017.

\bibitem{overstreet2022observation}
Chris Overstreet, Peter Asenbaum, Joseph Curti, Minjeong Kim, and Mark~A
  Kasevich.
\newblock Observation of a gravitational aharonov-bohm effect.
\newblock {\em Science}, 375(6577):226--229, 2022.

\bibitem{barrett2014sagnac}
Brynle Barrett, R{\'e}my Geiger, Indranil Dutta, Matthieu Meunier, Benjamin
  Canuel, Alexandre Gauguet, Philippe Bouyer, and Arnaud Landragin.
\newblock The sagnac effect: 20 years of development in matter-wave
  interferometry.
\newblock {\em Comptes Rendus Physique}, 15(10):875--883, 2014.

\bibitem{bertocchi2006single}
Guillaume Bertocchi, Olivier Alibart, Daniel~Barry Ostrowsky, S{\'e}bastien
  Tanzilli, and Pascal Baldi.
\newblock Single-photon sagnac interferometer.
\newblock {\em Journal of Physics B: Atomic, Molecular and Optical Physics},
  39(5):1011, 2006.

\bibitem{fink2017experimental}
Matthias Fink, Ana Rodriguez-Aramendia, Johannes Handsteiner, Abdul Ziarkash,
  Fabian Steinlechner, Thomas Scheidl, Ivette Fuentes, Jacques Pienaar,
  Timothy~C Ralph, and Rupert Ursin.
\newblock Experimental test of photonic entanglement in accelerated reference
  frames.
\newblock {\em Nature Communications}, 8(1):1--6, 2017.

\bibitem{restuccia2019photon}
Sara Restuccia, Marko Toro{\v{s}}, Graham~M Gibson, Hendrik Ulbricht, Daniele
  Faccio, and Miles~J Padgett.
\newblock Photon bunching in a rotating reference frame.
\newblock {\em Physical Review Letters}, 123(11):110401, 2019.

\bibitem{torovs2020revealing}
Marko Toro{\v{s}}, Sara Restuccia, Graham~M Gibson, Marion Cromb, Hendrik
  Ulbricht, Miles Padgett, and Daniele Faccio.
\newblock Revealing and concealing entanglement with noninertial motion.
\newblock {\em Physical {R}eview {A}}, 101(4):043837, 2020.

\bibitem{experiment}
Marion Cromb, Sara Restuccia, Graham~M Gibson, Marko Toro{\v{s}}, Miles~J
  Padgett, and Daniele Faccio.
\newblock Controlling photon entanglement with mechanical rotation.
\newblock {\em arXiv preprint arXiv:2210.05628}, 2022.

\bibitem{yin2017satellite}
Juan Yin, Yuan Cao, Yu-Huai Li, Sheng-Kai Liao, Liang Zhang, Ji-Gang Ren,
  Wen-Qi Cai, Wei-Yue Liu, Bo~Li, Hui Dai, et~al.
\newblock Satellite-based entanglement distribution over 1200 kilometers.
\newblock {\em Science}, 356(6343):1140--1144, 2017.

\bibitem{rivera2021outperforming}
Marco Rivera-Tapia, Marcel I~Y{\'a}{\~n}ez Reyes, A~Delgado, and G~Rubilar.
\newblock Outperforming classical estimation of post-newtonian parameters of
  earth's gravitational field using quantum metrology.
\newblock {\em arXiv:2101.12126}, 2021.

\bibitem{brady2021frame}
Anthony~J Brady and Stav Haldar.
\newblock Frame dragging and the hong-ou-mandel dip: Gravitational effects in
  multiphoton interference.
\newblock {\em Physical Review Research}, 3(2):023024, 2021.

\bibitem{kish2022quantum}
Sebastian~P Kish and Timothy~C Ralph.
\newblock Quantum effects in rotating reference frames.
\newblock {\em AVS Quantum Science}, 4(1):011401, 2022.

\bibitem{barzel2022observer}
Roy Barzel, David~Edward Bruschi, Andreas~W Schell, and Claus L{\"a}mmerzahl.
\newblock Observer dependence of photon bunching: The influence of the
  relativistic redshift on hong-ou-mandel interference.
\newblock {\em Physical Review D}, 105(10):105016, 2022.

\bibitem{mieling2022measuring}
Thomas Mieling, Christopher Hilweg, and Philip Walther.
\newblock Measuring space-time curvature using maximally path-entangled quantum
  states.
\newblock {\em arXiv:2202.12562}, 2022.

\bibitem{di2021sensitivity}
Angela~D Di~Virgilio, Carlo Altucci, Francesco Bajardi, Andrea Basti,
  Nicol{\`o} Beverini, Salvatore Capozziello, Giorgio Carelli, Donatella
  Ciampini, Francesco Fuso, Umberto Giacomelli, et~al.
\newblock Sensitivity limit investigation of a sagnac gyroscope through linear
  regression analysis.
\newblock {\em The European Physical Journal C}, 81(5):1--9, 2021.

\bibitem{di2017ginger}
Angela~DV Di~Virgilio, Jacopo Belfi, Wei-Tou Ni, Nicolo Beverini, Giorgio
  Carelli, Enrico Maccioni, and Alberto Porzio.
\newblock Ginger: A feasibility study.
\newblock {\em The European Physical Journal Plus}, 132(4):1--12, 2017.

\bibitem{capozziello2021constraining}
Salvatore Capozziello, Carlo Altucci, Francesco Bajardi, Andrea Basti,
  Nicol{\`o} Beverini, Giorgio Carelli, Donatella Ciampini, Angela~DV
  Di~Virgilio, Francesco Fuso, Umberto Giacomelli, et~al.
\newblock Constraining theories of gravity by ginger experiment.
\newblock {\em The European Physical Journal Plus}, 136(4):1--21, 2021.

\bibitem{everitt2011gravity}
CW~Francis Everitt, DB~DeBra, BW~Parkinson, JP~Turneaure, JW~Conklin,
  MI~Heifetz, GM~Keiser, AS~Silbergleit, T~Holmes, J~Kolodziejczak, et~al.
\newblock Gravity probe {B}: final results of a space experiment to test
  general relativity.
\newblock {\em Physical Review Letters}, 106(22):221101, 2011.

\bibitem{ciufolini2004confirmation}
Ignazio Ciufolini and Erricos~C Pavlis.
\newblock A confirmation of the general relativistic prediction of the
  lense--thirring effect.
\newblock {\em Nature}, 431(7011):958--960, 2004.

\bibitem{lacour2019sage}
Sylvester Lacour, FH~Vincent, M~Nowak, A~Le~Tiec, V~Lapeyrere, L~David,
  P~Bourget, A~Kellerer, K~Jani, J~Martino, et~al.
\newblock Sage: finding imbh in the black hole desert.
\newblock {\em Classical and Quantum Gravity}, 36(19):195005, 2019.

\bibitem{kim2019pulsed}
Heonoh Kim, Osung Kwon, and Han~Seb Moon.
\newblock Pulsed sagnac source of polarization-entangled photon pairs in
  telecommunication band.
\newblock {\em Scientific reports}, 9(1):1--7, 2019.

\bibitem{lee2021sagnac}
Youn~Seok Lee, Mengyu Xie, Ramy Tannous, and Thomas Jennewein.
\newblock Sagnac-type entangled photon source using only conventional
  polarization optics.
\newblock {\em Quantum Science and Technology}, 6(2):025004, 2021.

\bibitem{fuentes2005alice}
Ivette Fuentes-Schuller and Robert~B Mann.
\newblock Alice falls into a black hole: entanglement in noninertial frames.
\newblock {\em Physical {R}eview {L}etters}, 95(12):120404, 2005.

\bibitem{alsing2012observer}
Paul~M Alsing and Ivette Fuentes.
\newblock Observer-dependent entanglement.
\newblock {\em Classical and Quantum Gravity}, 29(22):224001, 2012.

\bibitem{sagnac1913ether}
Georges Sagnac.
\newblock L'{\'e}ther lumineux d{\'e}montr{\'e} par l'effet du vent relatif
  d'{\'e}ther dans un interf{\'e}rom{\`e}tre en rotation uniforme.
\newblock {\em CR Acad. Sci.}, 157:708--710, 1913.

\bibitem{sagnac1913preuve}
Georges Sagnac.
\newblock Sur la preuve de la r{\'e}alit{\'e} de l'{\'e}ther lumineux par
  l'exp{\'e}rience de l'interf{\'e}rographe tournant.
\newblock {\em CR Acad. Sci.}, 157:1410--1413, 1913.

\bibitem{einstein1909letter}
Albert Einstein.
\newblock Letter to {S}ommerfeld.
\newblock {\em Albert Einstein-Arnold Sommerfeld. Briefwechsel. Armin Hermann,
  ed. Basel and Stuttgart}, 1909.

\bibitem{korsbakken2004fulling}
Jan~Ivar Korsbakken and Jon~Magne Leinaas.
\newblock Fulling-{U}nruh effect in general stationary accelerated frames.
\newblock {\em Physical {R}eview {D}}, 70(8):084016, 2004.

\bibitem{padgett2008diffraction}
Miles~J Padgett.
\newblock On diffraction within a dielectric medium as an example of the
  {M}inkowski formulation of optical momentum.
\newblock {\em Optics {E}xpress}, 16(25):20864--20868, 2008.

\bibitem{barnett2010resolution}
Stephen~M Barnett.
\newblock Resolution of the abraham-minkowski dilemma.
\newblock {\em Physical Review Letters}, 104(7):070401, 2010.

\bibitem{post1967sagnac}
Evert~Jan Post.
\newblock Sagnac effect.
\newblock {\em Reviews of Modern Physics}, 39(2):475, 1967.

\bibitem{anderson1994sagnac}
Ronald Anderson, HR~Bilger, and GE~Stedman.
\newblock "{S}agnac" effect: A century of earth-rotated interferometers.
\newblock {\em American Journal of Physics}, 62(11):975--985, 1994.

\bibitem{malykin2000sagnac}
Grigorii~B Malykin.
\newblock The sagnac effect: correct and incorrect explanations.
\newblock {\em Physics-Uspekhi}, 43(12):1229, 2000.

\bibitem{thorne2000gravitation}
Kip~S Thorne, Charles~W Misner, and John~Archibald Wheeler.
\newblock {\em Gravitation}.
\newblock Freeman San Francisco, CA, 2000.

\bibitem{vallado2001fundamentals}
David~A Vallado.
\newblock {\em Fundamentals of astrodynamics and applications}, volume~12.
\newblock Springer Science \& Business Media, 2001.

\bibitem{aharonov1984quantum}
Yakir Aharonov and Tirzah Kaufherr.
\newblock Quantum frames of reference.
\newblock {\em Physical Review D}, 30(2):368, 1984.

\bibitem{giacomini2019relativistic}
Flaminia Giacomini, Esteban Castro-Ruiz, and {\v{C}}aslav Brukner.
\newblock Relativistic quantum reference frames: the operational meaning of
  spin.
\newblock {\em Physical review letters}, 123(9):090404, 2019.

\bibitem{kim2003single}
Yoon-Ho Kim.
\newblock Single-photon two-qubit entangled states: Preparation and
  measurement.
\newblock {\em Physical Review A}, 67(4):040301, 2003.

\bibitem{hill1997entanglement}
Scott Hill and William~K Wootters.
\newblock Entanglement of a pair of quantum bits.
\newblock {\em Physical review letters}, 78(26):5022, 1997.

\bibitem{englert2001universal}
Berthold-Georg Englert, Christian Kurtsiefer, and Harald Weinfurter.
\newblock Universal unitary gate for single-photon two-qubit states.
\newblock {\em Physical Review A}, 63(3):032303, 2001.

\bibitem{guhne2009entanglement}
Otfried G{\"u}hne and G{\'e}za T{\'o}th.
\newblock Entanglement detection.
\newblock {\em Physics Reports}, 474(1-6):1--75, 2009.

\bibitem{adhikari2010swapping}
S~Adhikari, AS~Majumdar, Dipankar Home, and AK~Pan.
\newblock Swapping path-spin intraparticle entanglement onto spin-spin
  interparticle entanglement.
\newblock {\em EPL (Europhysics Letters)}, 89(1):10005, 2010.

\bibitem{kumari2019swapping}
Asmita Kumari, Abhishek Ghosh, Mohit~Lal Bera, and AK~Pan.
\newblock Swapping intraphoton entanglement to interphoton entanglement using
  linear optical devices.
\newblock {\em Physical Review A}, 99(3):032118, 2019.

\bibitem{chen2016photon}
Zihao Chen, Yao Zhou, and Jung-Tsung Shen.
\newblock Photon antibunching and bunching in a ring-resonator waveguide
  quantum electrodynamics system.
\newblock {\em Optics letters}, 41(14):3313--3316, 2016.

\bibitem{chen2017dissipation}
Zihao Chen, Yao Zhou, and Jung-Tsung Shen.
\newblock Dissipation-induced photonic-correlation transition in waveguide-qed
  systems.
\newblock {\em Physical Review A}, 96(5):053805, 2017.

\bibitem{chen2021two}
Zihao Chen, Yao Zhou, Jung-Tsung Shen, Pei-Cheng Ku, and Duncan Steel.
\newblock Two-photon controlled-phase gates enabled by photonic dimers.
\newblock {\em Physical Review A}, 103(5):052610, 2021.

\bibitem{zych2012general}
Magdalena Zych, Fabio Costa, Igor Pikovski, Timothy~C Ralph, and {\v{C}}aslav
  Brukner.
\newblock General relativistic effects in quantum interference of photons.
\newblock {\em Classical and Quantum Gravity}, 29(22):224010, 2012.

\bibitem{bose2017spin}
Sougato Bose, Anupam Mazumdar, Gavin~W Morley, Hendrik Ulbricht, Marko
  Toro{\v{s}}, Mauro Paternostro, Andrew~A Geraci, Peter~F Barker, MS~Kim, and
  Gerard Milburn.
\newblock Spin entanglement witness for quantum gravity.
\newblock {\em Physical review letters}, 119(24):240401, 2017.

\bibitem{marletto2017gravitationally}
Chiara Marletto and Vlatko Vedral.
\newblock Gravitationally induced entanglement between two massive particles is
  sufficient evidence of quantum effects in gravity.
\newblock {\em Physical review letters}, 119(24):240402, 2017.

\bibitem{marshman2020locality}
Ryan~J Marshman, Anupam Mazumdar, and Sougato Bose.
\newblock Locality and entanglement in table-top testing of the quantum nature
  of linearized gravity.
\newblock {\em Physical Review A}, 101(5):052110, 2020.

\bibitem{danielson2022gravitationally}
Daine~L Danielson, Gautam Satishchandran, and Robert~M Wald.
\newblock Gravitationally mediated entanglement: Newtonian field versus
  gravitons.
\newblock {\em Physical Review D}, 105(8):086001, 2022.

\bibitem{bose2022mechanism}
Sougato Bose, Anupam Mazumdar, Martine Schut, and Marko
  Toro\ifmmode~\check{s}\else \v{s}\fi{}.
\newblock Mechanism for the quantum natured gravitons to entangle masses.
\newblock {\em Phys. Rev. D}, 105:106028, May 2022.

\bibitem{christodoulou2022locally}
Marios Christodoulou, Andrea Di~Biagio, Markus Aspelmeyer, {\v{C}}aslav
  Brukner, Carlo Rovelli, and Richard Howl.
\newblock Locally mediated entanglement through gravity from first principles.
\newblock {\em arXiv preprint arXiv:2202.03368}, 2022.

\bibitem{amelino2013quantum}
Giovanni Amelino-Camelia.
\newblock Quantum-spacetime phenomenology.
\newblock {\em Living Reviews in Relativity}, 16(1):1--137, 2013.

\bibitem{chou2017interferometric}
Aaron Chou, Henry Glass, H~Richard Gustafson, Craig~J Hogan, Brittany~L Kamai,
  Ohkyung Kwon, Robert Lanza, Lee McCuller, Stephan~S Meyer, Jonathan~W
  Richardson, et~al.
\newblock Interferometric constraints on quantum geometrical shear noise
  correlations.
\newblock {\em Classical and Quantum Gravity}, 34(16):165005, 2017.

\bibitem{vermeulen2021experiment}
Sander~M Vermeulen, Lorenzo Aiello, Aldo Ejlli, William~L Griffiths, Alasdair~L
  James, Katherine~L Dooley, and Hartmut Grote.
\newblock An experiment for observing quantum gravity phenomena using twin
  table-top 3d interferometers.
\newblock {\em Classical and Quantum Gravity}, 38(8):085008, 2021.

\bibitem{verlinde2021observational}
Erik~P Verlinde and Kathryn~M Zurek.
\newblock Observational signatures of quantum gravity in interferometers.
\newblock {\em Physics Letters B}, 822:136663, 2021.

\bibitem{pikovski2012probing}
Igor Pikovski, Michael~R Vanner, Markus Aspelmeyer, MS~Kim, and {\v{C}}aslav
  Brukner.
\newblock Probing {P}lanck-scale physics with quantum optics.
\newblock {\em Nature Physics}, 8(5):393--397, 2012.

\bibitem{amelino1998tests}
Giovanni Amelino-Camelia, John Ellis, NE~Mavromatos, Dimitri~V Nanopoulos, and
  Subir Sarkar.
\newblock Tests of quantum gravity from observations of $\gamma$-ray bursts.
\newblock {\em Nature}, 393(6687):763--765, 1998.

\bibitem{kostelecky2011data}
V~Alan Kosteleck{\`y} and Neil Russell.
\newblock Data tables for lorentz and c p t violation.
\newblock {\em Reviews of Modern Physics}, 83(1):11, 2011.

\bibitem{mattingly2005modern}
David Mattingly.
\newblock Modern tests of lorentz invariance.
\newblock {\em Living Reviews in relativity}, 8(1):1--84, 2005.

\bibitem{penrose1996gravity}
Roger Penrose.
\newblock On gravity's role in quantum state reduction.
\newblock {\em General relativity and gravitation}, 28(5):581--600, 1996.

\bibitem{donadi2021underground}
Sandro Donadi, Kristian Piscicchia, Catalina Curceanu, Lajos Di{\'o}si,
  Matthias Laubenstein, and Angelo Bassi.
\newblock Underground test of gravity-related wave function collapse.
\newblock {\em Nature Physics}, 17(1):74--78, 2021.

\bibitem{PhysRevLett.129.080401}
I.~J. Arnquist, F.~T. Avignone, A.~S. Barabash, C.~J. Barton, K.~H. Bhimani,
  E.~Blalock, B.~Bos, M.~Busch, M.~Buuck, T.~S. Caldwell, Y-D. Chan, C.~D.
  Christofferson, P.-H. Chu, M.~L. Clark, C.~Cuesta, J.~A. Detwiler, Yu.
  Efremenko, H.~Ejiri, S.~R. Elliott, G.~K. Giovanetti, M.~P. Green,
  J.~Gruszko, I.~S. Guinn, V.~E. Guiseppe, C.~R. Haufe, R.~Henning,
  D.~Hervas~Aguilar, E.~W. Hoppe, A.~Hostiuc, I.~Kim, R.~T. Kouzes, T.~E.
  Lannen~V., A.~Li, A.~M. Lopez, J.~M. L\'opez-Casta\~no, E.~L. Martin, R.~D.
  Martin, R.~Massarczyk, S.~J. Meijer, T.~K. Oli, G.~Othman, L.~S. Paudel,
  W.~Pettus, A.~W.~P. Poon, D.~C. Radford, A.~L. Reine, K.~Rielage, N.~W. Ruof,
  D.~Tedeschi, R.~L. Varner, S.~Vasilyev, J.~F. Wilkerson, C.~Wiseman, W.~Xu,
  C.-H. Yu, and B.~X. Zhu.
\newblock Search for spontaneous radiation from wave function collapse in the
  majorana demonstrator.
\newblock {\em Phys. Rev. Lett.}, 129:080401, Aug 2022.

\bibitem{bassi2017gravitational}
Angelo Bassi, Andr{\'e} Gro{\ss}ardt, and Hendrik Ulbricht.
\newblock Gravitational decoherence.
\newblock {\em Classical and Quantum Gravity}, 34(19):193002, 2017.

\bibitem{michelson1925effect1}
Albert~Abraham Michelson.
\newblock The {E}ffect of the {E}arth's {R}otation on the {V}elocity of
  {L}ight, i.
\newblock {\em The Astrophysical Journal}, 61:137, 1925.

\bibitem{michelson1925effect2}
Albert~Abraham Michelson and Henry~G Gale.
\newblock The {E}ffect of the {E}arth's {R}otation on the {V}elocity of
  {L}ight, ii.
\newblock {\em The Astrophysical Journal}, 61:140, 1925.

\bibitem{bachor2019guide}
Hans-A Bachor and Timothy~C Ralph.
\newblock {\em A guide to experiments in quantum optics}.
\newblock John Wiley \& Sons, 2019.

\end{thebibliography}

\clearpage{}
\begin{center}
\textbf{\large{} Supplemental material}{\large\par}
\par\end{center}

\setcounter{equation}{0}

\setcounter{section}{0} \setcounter{figure}{0} \setcounter{table}{0}
\makeatletter 
\global\long\def\theequation{S\arabic{equation}}%
\global\long\def\thefigure{S\arabic{figure}}%


\subsection{{Initial state preparation and basic definitions\label{state-preparation}}}

{We suppose we have the prepared the photon state:}

{
\begin{align}
\vert\psi_{\text{source}}\rangle & =(1,0,0,0)^{\top},\label{eq:source}
\end{align}
corresponding to the photon on a single path and polarization $\vert H\rangle$.
The state in Eq.~\eqref{eq:source} is a two-qubit state in $\mathcal{H}_{{\cal P}}\varotimes{\cal H}_{\Pi}$,
where ${\cal H}_{{\cal P}}$ (${\cal H}_{\Pi}$) denote the path (polarization)
Hilbert space. Such a state can be prepared using a photon
source and optical elements (see for example~\cite{bachor2019guide}).}
{Let us now see how the optical elements shown in Fig.~1(a) transform
the state in Eq.~\eqref{eq:source} to the initial state defined in
Eq.~(3).}

{We first recall the quarter-wave plate (QWP) and half-wave
plate (HWP) matrix transformations:}

{
\begin{alignat}{1}
U_{\text{QWP}} & (\theta)=\frac{1}{\sqrt{2}}\left[\begin{array}{cc}
1-i\cos(2\theta) & -i\sin(2\theta)\\
-i\sin(2\theta) & 1+i\cos(2\theta)
\end{array}\right],\label{eq:QWP}\\
U_{\text{HWP}} & (\theta)=-i\left[\begin{array}{cc}
\cos(2\theta) & \sin(2\theta)\\
\sin(2\theta) & -\cos(2\theta)
\end{array}\right],\label{eq:HWP}
\end{alignat}
respectively, where $\theta$ denotes the angle of rotation.}

{The first half-wave plate shown Fig.~1(a) is described
by the matrix transformation 
\begin{equation}
\mathcal{U_{\text{HWP1}}}=\left[\begin{array}{cc}
U_{\text{\text{HWP}}}(\pi/8) & 0\\
0 & 1
\end{array}\right].\label{eq:HWP1}
\end{equation}
We then construct the transformation matrix for a polarizing beam
splitter (PBS), where the basis vectors are given by:}

{
\begin{alignat}{1}
\vert aH\rangle & =\frac{1}{\sqrt{2}}(1,0,0,0)^{\top},\\
\vert aV\rangle & =\frac{1}{\sqrt{2}}(0,1,0,0)^{\top},\\
\vert bH\rangle & =\frac{1}{\sqrt{2}}(0,0,1,0)^{\top},\\
\vert bV\rangle & =\frac{1}{\sqrt{2}}(0,0,0,1)^{\top}.
\end{alignat}
We define the PBS transformation as}

{
\begin{alignat}{1}
\text{\ensuremath{U_{\text{PBS}}}}= & \vert aH\rangle\langle aH\vert+\vert bH\rangle\langle bH\vert\nonumber \\
 & +\vert aV\rangle\langle bV\vert+\vert bV\rangle\langle aV\vert,
\end{alignat}
which gives the following transformation matrix}

{
\begin{equation}
U_{\text{PBS}}=\frac{1}{2}\left[\begin{array}{cccc}
1 & 0 & 0 & 0\\
0 & 0 & 0 & 1\\
0 & 0 & 1 & 0\\
0 & 1 & 0 & 0
\end{array}\right].\label{eq:PBS}
\end{equation}
Finally, the two HWPs after the PBS corresponds to the following matrix
transformation 
\begin{equation}
\mathcal{U_{\text{HWP2}}}=\left[\begin{array}{cc}
U_{\text{\text{HWP}}}(\pi/8) & 0\\
0 & U_{\text{\text{HWP}}}(3\pi/8)
\end{array}\right].\label{eq:HWP2}
\end{equation}
}

{Applying the transformations in Eqs.~\eqref{eq:HWP1},
\eqref{eq:PBS} and \eqref{eq:HWP2} to the state in Eq.~\eqref{eq:source}
we readily find}

{
\begin{alignat}{1}
\vert\psi_{\text{initial}}\rangle & =\mathcal{U_{\text{HWP2}}}\;U_{\text{PBS}}\;\mathcal{U_{\text{HWP1}}}\vert\psi_{\text{source}}\rangle\\
 & =\frac{1}{2}(1,1,1,1)^{\top},
\end{alignat}
which is the state given in Eq.~(3), written in vector form.}

\subsection{Bell state basis\label{bell-state-measurement}}

The state obtained in Eq.~\eqref{eq:psifinal} can be written in
vector form as

\begin{align}
\vert\psi_{\text{final}}\rangle & =\frac{1}{2}(e^{-\frac{i\Omega r_{a}pt}{\hbar}},e^{+\frac{i\Omega r_{a}pt}{\hbar}},e^{-\frac{i\Omega r_{b}pt}{\hbar}},e^{+\frac{i\Omega r_{b}pt}{\hbar}})^{\top}.\label{eq:general}
\end{align}
The state $\vert\psi_{\text{final}}\rangle$ can become
maximally entangled when the rotational frequency $\Omega$ reaches
any odd multiple of $\Omega_{\text{Bell}}$ defined in Eq.~\eqref{eq:bell}.
Specifically, we obtain the Bell states in Eqs.~(\ref{eq:psifinal22})
and (\ref{eq:psifinal33}) which can be written in vector form as
\begin{alignat}{1}
\vert\psi_{1}\rangle & =\frac{1}{2}(e^{-i\frac{\pi}{2}\delta r_{a}},e^{i\frac{\pi}{2}\delta r_{a}},e^{-i\frac{\pi}{2}\delta r_{b}},e^{i\frac{\pi}{2}\delta r_{b}})^{\top},\label{eq:A1}\\
\vert\psi_{2}\rangle & =\frac{1}{2}(e^{-i\frac{3\pi}{2}\delta r_{a}},e^{i\frac{3\pi}{2}\delta r_{a}},e^{-i\frac{3\pi}{2}\delta r_{b}},e^{i\frac{3\pi}{2}\delta r_{b}})^{\top},\label{eq:A5}
\end{alignat}
with $\delta r_{j}=r_{j}/\Delta r~(j=a,b)$. Following the main text,
we choose for simplicity $r_{b}=2r_{a}$ (such that $\Delta r=r_{b}-r_{a}=r_{a}=r_{b}/2$
and $\delta r_{a}=\delta r_{b}/2=1$), which transforms the states
in Eqs.~(\ref{eq:A1}) and (\ref{eq:A5}) to 
\begin{alignat}{1}
\vert\psi_{1}\rangle & =\frac{1}{2}(-i,i,-1,-1)^{\top},\label{eq:psi1s}\\
\vert\psi_{2}\rangle & =\frac{1}{2}(i,-i,-1,-1)^{\top}.\label{eq:psi2s}
\end{alignat}
Furthermore, we introduce two additional states

\begin{alignat}{1}
\vert\psi_{3}\rangle & =\frac{1}{2}(-1,-1,-i,i)^{\top},\label{eq:psi3s}\\
\vert\psi_{4}\rangle & =\frac{1}{2}(-1,-1,i,-i)^{\top}.\label{eq:psi4s}
\end{alignat}
In this way the states $\vert\psi_{j}\rangle\,(j=1,..,4)$ form an
orthonormal basis of $\mathcal{H}_{{\cal P}}\varotimes{\cal H}_{\Pi}$.

We can readily verify that the states $\vert\psi_{j}\rangle\,(j=1,..,4)$
are maximally entangled by computing the concurrence~\citep{hill1997entanglement}
\begin{equation}
C=2\vert\mathfrak{a}_{1}\mathfrak{a}_{4}-\mathfrak{a}_{2}\mathfrak{a}_{3}\vert,
\end{equation}
where $\mathfrak{a}_{1}$,..,$\mathfrak{a}_{4}$ denote the components
of the vector in Eq.~\eqref{eq:general}. For the states $\vert\psi_{j}\rangle\,(j=1,..,4)$
we find the value $C=1$ which is the maximum possible value for any
entangled state on $\mathcal{H}_{{\cal P}}\varotimes{\cal H}_{\Pi}$.

\subsection{Bell state projection scheme\label{subsec:Bell-state-projection}}

The Bell states defined in Eqs.~\eqref{eq:psi1s}-\eqref{eq:psi4s}
can be measured using the scheme shown in Fig.~1(b) \citep{kim2003single}.
The scheme consists of a polarizing beam splitter (PBS) at $45^{\circ}$
and$\text{ quarter-wave plates (QWPs) at angle 4\ensuremath{5^{\circ}}}$.

We first construct the transformation matrix for a polarizing beam
splitters (PBS) at $45^{\circ}$. The basis vectors are given by:

\begin{alignat}{1}
\vert aD\rangle & =\frac{1}{\sqrt{2}}(1,1,0,0)^{\top},\\
\vert aA\rangle & =\frac{1}{\sqrt{2}}(1,-1,0,0)^{\top},\\
\vert bD\rangle & =\frac{1}{\sqrt{2}}(0,0,1,1)^{\top},\\
\vert bA\rangle & =\frac{1}{\sqrt{2}}(0,0,1,-1)^{\top}.
\end{alignat}
We can then define the transformation of the PBS at $45^{\circ}$
as

\begin{alignat}{1}
U_{\text{PBS}}(\pi/4)= & \vert aD\rangle\langle aD\vert+\vert bD\rangle\langle bD\vert\nonumber \\
 & +\vert aA\rangle\langle bA\vert+\vert bA\rangle\langle aA\vert,
\end{alignat}
which gives the following transformation matrix

\begin{equation}
U_{\text{PBS}}(\pi/4)=\frac{1}{2}\left[\begin{array}{cccc}
1 & 1 & 1 & -1\\
1 & 1 & -1 & 1\\
1 & -1 & 1 & 1\\
-1 & 1 & 1 & 1
\end{array}\right].
\end{equation}
The QWP at angle $45^{\circ}$ is defined by the following transformation
matrix (see Eq.~\eqref{eq:QWP} for the general definition) 
\begin{equation}
U_{\text{\text{QWP}}}(\pi/4)\equiv\frac{e^{-\frac{i\pi}{4}}}{2}\left(\begin{array}{cc}
1+i & 1-i\\
1-i & 1+i
\end{array}\right).\label{eq:uqwp45}
\end{equation}
We can now define the transformation corrsponding to the scheme in
Fig.~1(b) as

\begin{equation}
\mathcal{U_{B}}=\left[\begin{array}{cc}
U_{\text{\text{QWP}}}(\pi/4) & 0\\
0 & U_{\text{\text{QWP}}}(\pi/4)
\end{array}\right]U_{\text{PBS}}(\pi/4).
\end{equation}
We find

\begin{alignat}{1}
\mathcal{U_{B}}\vert\psi_{1}\rangle & =-e^{-\frac{i\pi}{4}}(0,0,0,1)^{\top},\\
\mathcal{U_{B}}\vert\psi_{2}\rangle & =-e^{-\frac{i\pi}{4}}(0,0,1,0)^{\top},\\
\mathcal{U_{B}}\vert\psi_{3}\rangle & =-e^{-\frac{i\pi}{4}}(0,1,0,0)^{\top},\\
\mathcal{U_{B}}\vert\psi_{4}\rangle & =-e^{-\frac{i\pi}{4}}(1,0,0,0)^{\top},
\end{alignat}
where the global phase factors are not important here. The four detectors
$B_{1},..,B_{4}$ in Fig.~1(b) will thus give a maximum signal whenever
we are in one of the Bell state $\vert\psi_{1}\rangle,..,\vert\psi_{4}\rangle$,
respectively, while the other three detectors will show a null signal.
In our specific experimental configuration we see that the detectors
$B_{1}$ and $B_{2}$ can thus identify the two Bell states $\vert\psi_{1}\rangle$
and $\vert\psi_{2}\rangle$, respectively (see Fig.~1(c)).

\subsection{Construction of optimal entanglement witness\label{sec:ConstructionWitness}}

To assert the generated entanglement using the scheme in Fig.~\ref{fig:experiment},
we construct an optimal entanglement witness~\citep{guhne2009entanglement}.

We consider first the Bell state $\vert\psi_{1}\rangle$ defined in
Eq.~\eqref{eq:psifinal1S}, where we have assumed for simplicity
$\Delta r=r_{b}-r_{a}=r_{a}=r_{b}/2$ and $\delta r_{a}=\delta r_{b}/2=1$
following the main text. In order to estimate the robustness of the
witness that we are going to identify, we introduce the Werner state
\begin{equation}
\rho[p]=p\vert\psi_{1}\rangle\langle\psi_{1}\vert+\frac{(1-p)}{4}\mathbb{I}_{4},\label{eq:A2}
\end{equation}
where $p$ is the probability of preparing the desired state and $\mathbb{I}_{n}$
is a $n\times n$ identity matrix. The first (second) term on the
right-hand side of Eq.~(\ref{eq:A2}) denotes the contributions from
the ideal entangled state (a mixed non-entangled state which could
arise from any number of noise or decoherence sources and is typically
referred to as \textit{white noise}). Provided the probability of
preparing the desired states satisfies the condition $p>1/3$, the
partially transposed density matrix 
would have the negative eigenvalue $(1-3p)/4$, with the associated
eigenstate $\vert\phi_{\text{m}}\rangle=\frac{1}{2}(i,i,-1,1)^{\top},$
thus affirming entanglement in light of the Peres-Horodecki criterion.
One can then readily construct the optimal fidelity-based entanglement
witness $\mathcal{W}\equiv\vert\phi_{\text{m}}\rangle\langle\phi_{\text{m}}\vert^{T_{\Pi}}$,
which reduces to 
\begin{equation}
\mathcal{W}_{1}=\frac{1}{4}\left[\begin{array}{cccc}
1 & 1 & -i & -i\\
1 & 1 & i & i\\
i & -i & 1 & -1\\
i & -i & -1 & 1
\end{array}\right].\label{eq:A3}
\end{equation}
One can then show that the witness in Eq.~(\ref{eq:A3}) can be decomposed
in the following set of local operations 
\begin{equation}
\mathcal{W}_{1}=\frac{1}{4}\left[\mathbb{I}_{4}+\sigma_{x}^{{\cal P}}\sigma_{y}^{\Pi}+\sigma_{y}^{{\cal P}}\sigma_{z}^{\Pi}+\sigma_{z}^{{\cal P}}\sigma_{x}^{\Pi}\right],\label{eq:W1}
\end{equation}
where $\sigma_{k}^{{\cal J}}$ denotes the $k=x,y,z$ Pauli operator
for ${\cal J}={\cal P}$, ${\Pi}$.

A similar calculation can be performed for the other maximally entangled
state $\vert\psi_{2}\rangle$ in Eq.~\eqref{eq:psi2s}. Specifically,
following the steps in Eqs.~(\ref{eq:A1})-(\ref{eq:A3}) we find
the witness 
\begin{equation}
\mathcal{W}_{2}=\frac{1}{4}\left[\mathbb{I}_{4}-\sigma_{x}^{{\cal P}}\sigma_{y}^{\Pi}-\sigma_{y}^{{\cal P}}\sigma_{z}^{\Pi}+\sigma_{z}^{{\cal P}}\sigma_{x}^{\Pi}\right].\label{eq:W2}
\end{equation}
Therefore, the set of local operations needed to reconstruct the value
of $\mathcal{W}_{1}$ and $\mathcal{W}_{2}$ is the same.

\subsection{Measuring entanglement witnesses using the universal unitary gate\label{universal-unitary-gate} }

The entanglement witnesses in Eqs.~\eqref{eq:W1} and \eqref{eq:W2}
can be implemented using the universal unitary gate for single-photon
two-qubit states shown in Fig.~1(d)~\citep{englert2001universal}.
Such a gate can be used to perform a measurement of a generic 2-qubit
observable $\mathcal{O}$ using linear optical elements.

\begin{table*}[t]
\begin{tabular}{|c||c|c|c|c|}
\hline 
$\mathcal{O}$  & $S_{\text{RR}}$  & $S_{\text{LL}}$  & $S_{\text{RL}}$  & $S_{\text{LR}}$\tabularnewline
\hline 
\hline 
$\sigma_{x}^{{\cal P}}\sigma_{y}^{\Pi}$  & $\frac{1}{\sqrt{2}}\sigma_{z}^{\Pi}$  & $\frac{1}{\sqrt{2}}\sigma_{x}^{\Pi}$  & $-\frac{i}{\sqrt{2}}\sigma_{x}^{\Pi}$  & $-\frac{i}{\sqrt{2}}\sigma_{z}^{\Pi}$\tabularnewline
\hline 
$\sigma_{y}^{{\cal P}}\sigma_{z}^{\Pi}$  & $\frac{1}{\sqrt{2}}\sigma_{x}^{\Pi}$  & $\frac{1+i}{2\sqrt{2}}(\sigma_{x}^{\Pi}+\sigma_{y}^{\Pi})$  & $-\frac{1}{\sqrt{2}}\sigma_{y}^{\Pi}$  & $\frac{1+i}{2\sqrt{2}}(\sigma_{x}^{\Pi}-\sigma_{y}^{\Pi})$\tabularnewline
\hline 
$\sigma_{z}^{{\cal P}}\sigma_{x}^{\Pi}$  & $\frac{1}{2\sqrt{2}}(-\mathbb{I}^{\Pi}+\sigma_{x}^{\Pi}-i\sigma_{y}^{\Pi}+\sigma_{z}^{\Pi})$  & $\frac{1}{2\sqrt{2}}(\mathbb{I}^{\Pi}-\sigma_{x}^{\Pi}-i\sigma_{y}^{\Pi}+\sigma_{z}^{\Pi})$  & $\frac{1}{2\sqrt{2}}(\mathbb{I}^{\Pi}+\sigma_{x}^{\Pi}+i\sigma_{y}^{\Pi}+\sigma_{z}^{\Pi})$  & $\frac{1}{2\sqrt{2}}(\mathbb{I}^{\Pi}+\sigma_{x}^{\Pi}-i\sigma_{y}^{\Pi}-\sigma_{z}^{\Pi})$\tabularnewline
\hline 
\end{tabular}

\caption{Components of the unitary 2-qubit gate defined in Eq.~\eqref{eq:Smatrix}
which maps the four eigenstates of the 2-qubit observable $\mathcal{\mathbb{\mathcal{O}}}$
to the four output ports $D_{j}$ ($j=1,..,4$) shown in Fig.~1(d)
.\label{table1}}

\begin{tabular}{|c||c|c|c|c|c|c|c|c|c|c|c|c|c|c|c|c|c|c|c|c|}
\hline 
$\mathcal{O}$  & $V_{1}$  & $\alpha$  & $\beta$  & $\gamma$  & $\delta$  & $V_{L}$  & $\alpha$  & $\beta$  & $\gamma$  & $\delta$  & $V_{R}$  & $\alpha$  & $\beta$  & $\gamma$  & $\delta$  & $V_{2}$  & $\alpha$  & $\beta$  & $\gamma$  & $\delta$\tabularnewline
\hline 
\hline 
$\sigma_{x}^{{\cal P}}\sigma_{y}^{\Pi}$  & $-\sigma_{y}^{\Pi}$  & $0$  & $\frac{\pi}{4}$  & $\frac{\pi}{2}$  & $\frac{\pi}{2}$  & $e^{i\frac{\pi}{4}}\sigma_{x}^{\Pi}$  & $0$  & $\frac{\pi}{4}$  & $0$  & $\frac{3\pi}{4}$  & $e^{-i\frac{\pi}{4}}\sigma_{x}^{\Pi}$  & $\frac{\pi}{4}$  & $-\frac{\pi}{4}$  & $-\frac{\pi}{4}$  & $-\frac{3\pi}{4}$  & $i\mathbb{I}^{\Pi}$  & $\frac{\pi}{4}$  & $0$  & $\frac{\pi}{4}$  & $\pi$\tabularnewline
\hline 
$\sigma_{y}^{{\cal P}}\sigma_{z}^{\Pi}$  & $-i\sigma_{z}^{\Pi}$  & $0$  & $0$  & $\frac{\pi}{2}$  & $0$  & $\frac{i}{\sqrt{2}}(\mathbb{\sigma}_{x}^{\Pi}+\sigma_{y}^{\Pi})$  & $-\frac{\pi}{4}$  & $\frac{3\pi}{8}$  & $\frac{\pi}{4}$  & $-\pi$  & $\frac{1}{\sqrt{2}}(\mathbb{\sigma}_{x}^{\Pi}+\sigma_{y}^{\Pi})$  & $-\frac{\pi}{4}$  & $-\frac{\pi}{8}$  & $\frac{\pi}{4}$  & $-\frac{\pi}{2}$  & $\frac{1-i}{2}\mathbb{I}^{\Pi}-\frac{1+i}{2}\sigma_{z}^{\Pi}$  & $-\frac{\pi}{4}$  & $\frac{\pi}{8}$  & $-\frac{\pi}{4}$  & $-\frac{\pi}{4}$\tabularnewline
\hline 
$\sigma_{z}^{{\cal P}}\sigma_{x}^{\Pi}$  & $-i\sigma_{z}^{\Pi}$  & $0$  & $0$  & $0$  & $\frac{\pi}{2}$  & $\frac{1}{\sqrt{2}}(\sigma_{z}^{\Pi}-\sigma_{x}^{\Pi})$  & $0$  & $-\frac{\pi}{8}$  & $\frac{\pi}{4}$  & $\frac{\pi}{2}$  & $\frac{1}{\sqrt{2}}(\mathbb{I}^{\Pi}-i\sigma_{y}^{\Pi})$  & $0$  & $\frac{\pi}{8}$  & $\frac{\pi}{4}$  & $\pi$  & $i\sigma_{x}^{\Pi}$  & $0$  & $-\frac{\pi}{4}$  & $0$  & $0$\tabularnewline
\hline 
\end{tabular}

\caption{Configuration of the angles of the HWP, QWPs, and phase shifter to
measure the entanglement witnesses defined in Eqs.~\eqref{eq:W1}
and \eqref{eq:W2}.\label{table2}}
\end{table*}

\subsubsection{Constructing the gate}

We first summarize how to construct the unitary gate given a 2-qubit
observable $\mathcal{O}$. We first find the eigenvectors of the observable
$\mathcal{O}$ which we denote by $\vert e_{j}\rangle$ ($j=1,..,4$)
and then construct the operators

\begin{alignat}{1}
A & =\vert e_{1}\rangle\langle e_{1}\vert+\vert e_{2}\rangle\langle e_{2}\vert-\vert e_{3}\rangle\langle e_{3}\vert-\vert e_{4}\rangle\langle e_{4}\vert,\\
B & =\vert e_{1}\rangle\langle e_{1}\vert-\vert e_{2}\rangle\langle e_{2}\vert+\vert e_{3}\rangle\langle e_{3}\vert-\vert e_{4}\rangle\langle e_{4}\vert.
\end{alignat}
The gate $S$ is then defined as the transformations that achieves

\begin{equation}
SAS^{\dagger}=\sigma_{z}^{{\cal P}},\qquad SBS^{\dagger}=\sigma_{z}^{\Pi}.
\end{equation}
The measurement of the path observable $\sigma_{z}^{{\cal P}}$ distinguishes
between $\vert e_{1}\rangle,\vert e_{2}\rangle$ (eigenvalue $+1$)
and $\vert e_{3}\rangle,\vert e_{4}\rangle$ (eigenvalue $-1$), while
the measurement of the polarization observable $\sigma_{z}^{\Pi}$
selects beween $\vert e_{1}\rangle,\vert e_{3}\rangle$ (eigenvalue
$+1$) and $\vert e_{2}\rangle,\vert e_{4}\rangle$ (eigenvalue $-1$)
-- the two meausurements together thus uniqely identify the eigenstate.
Hence the transfromation $S$ maps eigenstates of $\mathcal{O}$ to
the output ports where they are measured by the detectors $D_{j}$
($j=1,..,4$) shown in Fig.~1(d). In this sense the universal unitary
gate can be seen as a generalization of the (simpler) Bell state projection
gate discussed in Sec.~\ref{subsec:Bell-state-projection}.

The gate $S$ can be conveniently written in component form as

\begin{equation}
S=\left[\begin{array}{cc}
S_{\text{RR}} & S_{\text{RL}}\\
S_{\text{LR}} & S_{\text{LL}}
\end{array}\right],\label{eq:Smatrix}
\end{equation}
where the $2\times2$ matrix acts on ${\cal H}^{\mathcal{P}}$, and
the matrices $S_{\text{RR}},S_{\text{LL}},S_{\text{RL}},S_{\text{LR}}$
on the polarization Hilbert space ${\cal H}^{\Pi}$. We recall that
the entanglement witnesses $\mathcal{W}_{1}$ and $\mathcal{W}_{2}$
defined in Eqs.~\eqref{eq:W1} and \eqref{eq:W2}) are composed by
three observables: $\sigma_{x}^{{\cal P}}\sigma_{y}^{\Pi},\sigma_{y}^{{\cal P}}\sigma_{z}^{\Pi}$
and $\sigma_{z}^{{\cal P}}\sigma_{x}^{\Pi}$ . We list the components
of the matrix $S$ for each of these observables in Table.~\ref{table1}.

\subsubsection{Implementing the gate}

The universal gate shown Fig.~1(d) consists of two beam-splitters
(BSs), which together with the mirrors forms a Mach-Zehnder interferometer,
and of four composite optical elements denoted by $V_{j}$ ($j=1,2,R,L$).
Each element $V_{j}$ is composed of a half-wave plate (HWP) inserted
between two quarter-wave plates (QWPs) and an additional phase shifter.
The associated unitary transformation is given by

\begin{equation}
V_{j}=e^{i\delta}U_{\text{QWP}}(\gamma)U_{\text{HWP}}(\beta)U_{\text{QWP}}(\alpha),\label{eq:Vj}
\end{equation}
where $\delta$ is a phase shift, and $\alpha$, $\beta$, $\gamma$
denote the first QWP, middle HWP, and second QWP angle, respectively.
The QWP and HWP transformations matrices are given in Eqs.~\eqref{eq:QWP}
and \eqref{eq:HWP}, respectively.

Let us summarize the procedure for choosing the angles $\alpha$,
$\beta$, $\gamma,\delta$ that will realize the gates $S$ obtained
in Table.~\ref{table1} (and hence the measurement of the corresponding
observables $\mathcal{O}$). The first step is to decompose the matrix
$S$ exploiting its unitary character. In particular, it has been
shown that the matrix S can be rewritten in the following way:

\begin{alignat}{1}
S_{RR} & =\text{cos}(\theta_{1})\vert\bar{\psi}_{1}\rangle\langle\psi_{1}\vert+\text{cos}(\theta_{2})\vert\bar{\psi}_{2}\rangle\langle\psi_{2}\vert,\label{eq:SRR}\\
S_{LL} & =\text{cos}(\theta_{1})\vert\bar{\chi}_{1}\rangle\langle\chi_{1}\vert+\text{cos}(\theta_{2})\vert\bar{\chi}_{2}\rangle\langle\chi_{2}\vert,\\
S_{RL} & =-i(\text{sin}(\theta_{1})\vert\bar{\psi}_{1}\rangle\langle\chi_{1}\vert+\text{sin}(\theta_{2})\vert\bar{\psi}_{2}\rangle\langle\chi_{2}\vert),\\
S_{LR} & =-i(\text{sin}(\theta_{1})\vert\bar{\chi}_{1}\rangle\langle\psi_{1}\vert+\text{sin}(\theta_{2})\vert\bar{\chi}_{2}\rangle\langle\psi_{2}\vert).\label{eq:SLR}
\end{alignat}
The left-hand side of Eqs.~\eqref{eq:SRR}-\eqref{eq:SLR} are listed
in Table~\ref{table1} and we can determine the angles and vectors
appearing on the right-hand side in the following way. From $S_{RR}^{\dagger}S_{RR}$
we first obtain the eigenvectors $\vert\psi_{j}\rangle$ ($j=1,2$)
as well as the angles $\theta_{j}$ ($j=1,2$), up to arbitary phases.
From $S_{RR}S_{RR}^{\dagger}$,$S_{LL}^{\dagger}S_{LL}$, and $S_{LL}S_{LL}^{\dagger}$
we then determine the eigenvectors $\vert\bar{\psi}_{j}\rangle$,$\vert\chi_{j}\rangle$,
and $\vert\bar{\chi}_{j}\rangle$, respectively. Using the decomposition
in Eqs.~\eqref{eq:SRR}-\eqref{eq:SLR} we then construct the four
transformation matrices:

\begin{alignat}{1}
V_{1} & =-i\vert\chi_{1}\rangle\langle\psi_{1}\vert-i\vert\chi_{2}\rangle\langle\psi_{2}\vert,\label{eq:V1}\\
V_{2} & =i\vert\bar{\psi}_{1}\rangle\langle\bar{\chi}_{1}\vert+i\vert\bar{\psi}_{2}\rangle\langle\bar{\chi}_{2}\vert,\\
V_{R} & =e^{-i\theta_{1}}\vert\bar{\chi}_{1}\rangle\langle\chi_{1}\vert+e^{-i\theta_{2}}\vert\bar{\chi}_{2}\rangle\langle\chi_{2}\vert,\\
V_{L} & =e^{i\theta_{1}}\vert\bar{\chi}_{1}\rangle\langle\chi_{1}\vert+e^{i\theta_{2}}\vert\bar{\chi}_{2}\rangle\langle\chi_{2}\vert.\label{eq:V2}
\end{alignat}
The matrices $V_{j}$ ($j=1,2,R,L$) are then finally decomposed in
the form of Eq.~\ref{eq:Vj} by a suitable choice of the angles $\alpha$,
$\beta$, $\gamma,\delta$. The matrices $V_{j}$ ($j=1,2,R,L$) together
with the angles $\alpha$, $\beta$, $\gamma,\delta$ are listed in
Table.~\ref{table2}.

\subsection{Alternative single loop scheme}

\begin{figure*}[t]
\includegraphics[width=1\textwidth]{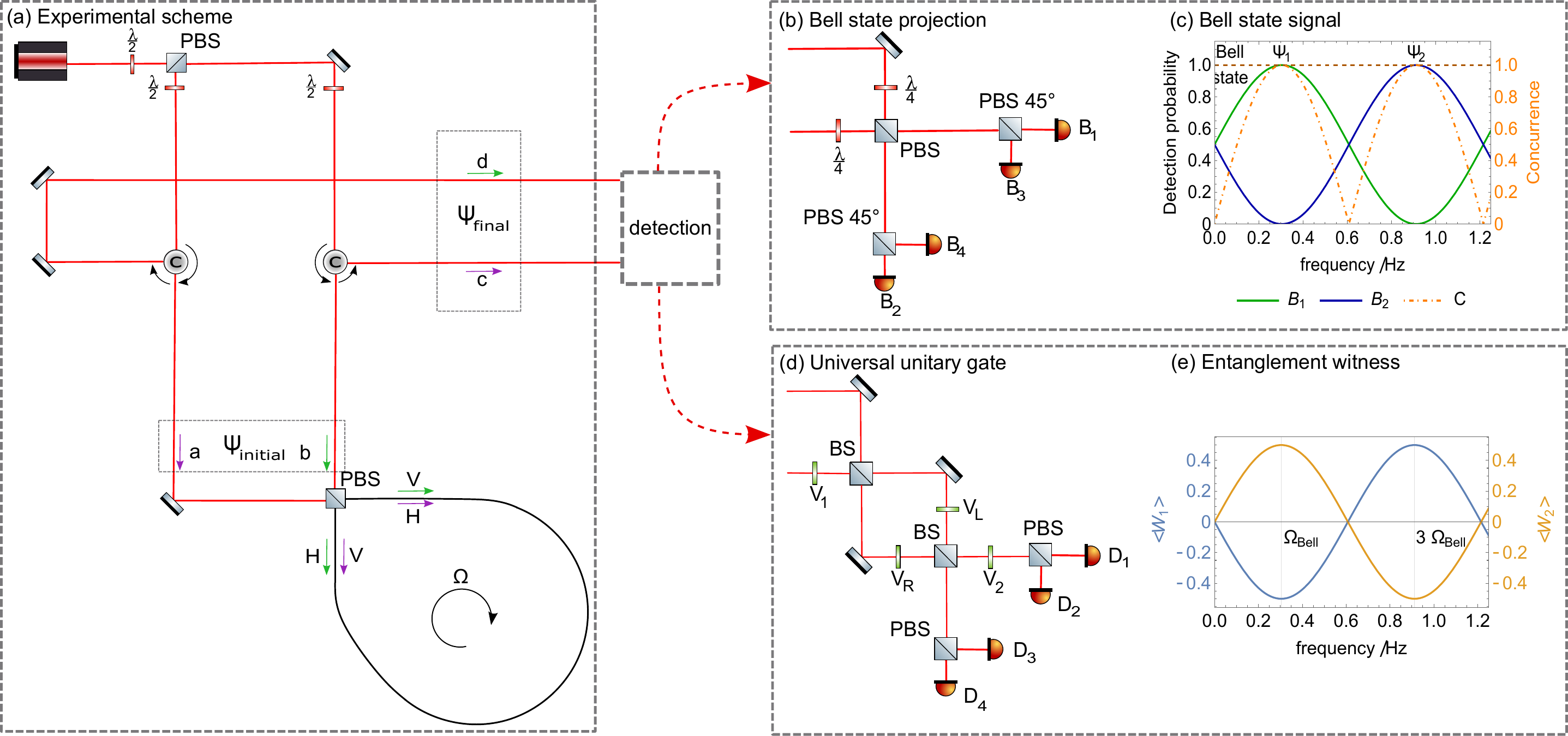}

\caption{Single loop scheme for generating path-polarization entanglement from
mechanical rotation. \textbf{(a) }The scheme consists of a single
photon source, polarizing beam splitters (PBSs), and half-wave plates
(HWPs) denoted by $\lambda/2$, and two optical circulators denoted by the
letter C in a circle. We consider the same experimental values as
in Fig.~1 of the main text (with the radius of the single loop set
to $r=0.5\text{m}$). The experimental setup is placed on a platform
which can be set in rotation with frequency $\Omega$. The purple
and green arrows indicate the paths $a$, $b$, while the polarization
is denoted by $H$, $V$. We start with a separable path-polarization
photon $\vert\psi_{\text{initial}}\rangle$ -- depending on the frequency
of rotation $\Omega$ the final photon $\vert\psi_{\text{final}}\rangle$
can remain separable or becomes entangled. We can generate maximally
entangled Bell states by tuning the frequency of the platform to $\Omega_{\text{Bell}}=\mathcal{N}\pi c^{2}/(8\omega\mathcal{A})$,
where $\omega$ is the mean photon frequency, and $\mathcal{A}$ is
the effective area of the interferometer (the path length assumed
to be equal for the $a$, $b$ paths), and $\mathcal{N}=1,3,5,...$.\textbf{
(b)} The detector $B_{j}$ measures the Bell-state $\psi_{j}$ ($j=1,..,4$).
The first PBS is rotated by $\pi/4$, and the quarter-wave plates
(QWPs) are denoted by $\lambda/4$ (with the fast axis oriented at
$\pi/4$ which transforms circular polarization to linear polarization).
\textbf{(c) }The detector $B1$ ($B2$) measures the maximally entangled
Bell-state $\psi_{1}$($\psi_{2}$) at frequency $\Omega_{\text{Bell}}$($3\Omega_{\text{Bell}}$)
where the concurrence achieves the maximum possible value $C=1$.
We find $\Omega_{\text{Bell}}\sim2\pi\times0.3\text{Hz}$, which can
be readily achieved with similar photonic setups~\citep{restuccia2019photon}.
\textbf{(d)} Universal unitary gate for single-photon two-qubit states
which can be used for tomographic reconstruction or implementing entanglement
witnesses. The two beam-splitters (BSs) together with the mirrors
form a Mach-Zehnder interferometer, and $V_{j}$($j=1,R,L,2$) denotes
an optical element composed of a HWP, two enclosing QWPs, and a phase
shifter. \textbf{(e)} Optimal entanglement witnesses $\mathcal{W}_{1}$
($\mathcal{W}_{2}$) for the maximally entangled Bell-state $\psi_{1}$($\psi_{2}$)
as a function of rotation frequency. Entanglement is witnesses when
$\mathcal{W}_{j}<0$ ($j=1,2$).}
\label{fig2} 
\end{figure*}

\label{singleloop}

We present an alternative implementation of the platform illustrated
in the main text of the paper. Specifically, we demonstrate that entangled
path-polarization states akin to those discussed in the main text
of the paper can be readily generated through a single-loop scheme.
To this end, we send the photon into four different paths as shown
in Fig.~\ref{fig2}. The corresponding four states $\vert a\rangle\vert H\rangle$,
$\vert a\rangle\vert V\rangle$,$\vert b\rangle\vert H\rangle$,$\vert b\rangle\vert V\rangle$
acquire Sagnac phases. Specifically, from the coupling $\sim rp$
in Eq.~(\ref{eq:Hamiltonian}) we find that the phases are given
by $\phi_{\pm}=\pm\Omega rpt/\hbar$, where $t=\frac{nl}{c}$ is the
the flight time, $n$ is the refractive index of the fiber, $l=2\pi r$
is the length of the fibers (automatically equal for the two paths),
$c$ is the speed of light, $p=E/(nc)=\hbar\omega/(nc)$ is the momentum,
and $\omega$ is the mean photon frequency. We note that the phases
can be rewritten as $\phi_{\pm}=\pm\Omega rpt/\hbar=\pm2\Omega\omega\mathcal{A}/c^{2}$,
where $\mathcal{A}=\pi r^{2}$ is the effective area of the interferometer.
The phases $\phi_{\pm}=\pm\phi_{s}/2$ can be seen as variants of
the Sagnac phase $\phi_{s}\equiv4\Omega\omega\mathcal{A}/c^{2}$ .
Importantly, the phases $\phi_{\pm}$ do not depend on the refractive
index of the medium which indicates that the effect is ultimately
linked to the geometry of spacetime~\citep{torovs2020revealing}.

The initial (non-entangled) state is given by

\noindent 
\begin{equation}
\vert\psi_{\text{initial}}\rangle=\frac{1}{2}(\vert a\rangle+\vert b\rangle)(\vert H\rangle+\vert V\rangle),\label{eq:initial2}
\end{equation}
and the final state is given by 
\begin{equation}
\begin{aligned}\vert\psi_{\text{final}}\rangle & =\frac{1}{2}\left(\vert a\rangle[e^{-i\frac{\phi_{s}}{2}}\vert H\rangle+e^{+i\frac{\phi_{s}}{2}}\vert V\rangle]\right.\\
 & +\left.\vert b\rangle[e^{+i\frac{\phi_{s}}{2}}\vert H\rangle+e^{-i\frac{\phi_{s}}{2}}\vert V\rangle]\right)
\end{aligned}
,\label{eq:psifinal2}
\end{equation}
where we have used the Hamiltonian $H_{\text{rot}}$ defined in Eq.~\eqref{eq:Hamiltonian}.

Let us consider first the case $\Omega=0$. The final state in Eq.~(\ref{eq:psifinal2})
reduces to the initial state in Eq.~(\ref{eq:initial2}) and thus
a non-rotating platform has no effect on entanglement. In contrast,
when $\Omega\neq0$ the state in Eq.~(\ref{eq:psifinal2}) will in
general become entangled (as we can no longer write it as the product
of the path and polarization states). This shows that non-inertial
rotating motion generates entanglement.

Along the same lines as those presented in the main body of the paper,
we look for the condition to achieve maximum entanglement, which is
achieved when 
the polarization states in the square brackets of Eq.~(\ref{eq:psifinal2})
become orthogonal. Such condition is witnessed by the null overlap
$S$ of the polarization states. 
We readily find 
$S=\text{cos}(\phi_{s})$, so that 
the rotation frequency should achieve the value 
\begin{equation}
\Omega_{\text{Bell}}\equiv\frac{(2k+1)\pi c^{2}}{8\omega\mathcal{A}}~~~(k\in\mathbb{Z}).
\end{equation}
However, if we set the rotation frequency $\Omega$ to any odd multiple
frequency of $\Omega_{\text{Bell}}$ we will also generate a maximally
entangled Bell state. On the other hand, the resulting state would
be separable for 
$\Omega=\pm2n\Omega_{\text{Bell}}~(n\in\mathbb{Z})$. 
Therefore, similar considerations to those made in the main body of
the paper in regard to the double-loop scheme can be put forward in
this case as well. 
In particular, the scheme will generate 
the Bell-states 
\begin{alignat}{1}
\vert\psi_{1}\rangle\sim & \vert a\rangle[(1-i)\vert H\rangle+(1+i)\vert V\rangle]\nonumber \\
 & +\vert b\rangle[(1+i)\vert H\rangle+(1-i)\vert V\rangle]\label{eq:psifinal1S}
\end{alignat}
at $\Omega/\Omega_{\text{Bell}}=...,-7,-3,1,5...$, and 
\begin{alignat}{1}
\vert\psi_{2}\rangle\sim & \vert a\rangle[(-1-i)\vert H\rangle+(-1+i)\vert V\rangle]\nonumber \\
 & +\vert b\rangle[(-1+i)\vert H\rangle+(-1-i)\vert V\label{eq:psifinal2S}
\end{alignat}
at $\Omega/\Omega_{\text{Bell}}=...,-5,-1,3,7...$. 

We summarize below the details of the two detections schemes shown
in Fig.~\ref{fig2}(b)-\ref{fig2}(e) following the analysis performed
for the double loop scheme in the Supplementary material \ref{bell-state-measurement}-\ref{universal-unitary-gate}.

\subsubsection{Bell state basis}

We recast the two states in Eqs.~\eqref{eq:psifinal1S} and \eqref{eq:psifinal2S}
in vector form:

\begin{alignat}{1}
\vert\psi_{1}\rangle & =\frac{1}{2\sqrt{2}}(1-i,1+i,1+1,1-1)^{\top},\label{eq:psi1s-1}\\
\vert\psi_{2}\rangle & =\frac{1}{2\sqrt{2}}(1-i,-1+i,-1+i,-1-i)^{\top}.\label{eq:psi2s-1}
\end{alignat}
Furthermore, we introduce two additional states

\begin{alignat}{1}
\vert\psi_{3}\rangle & =\frac{1}{2\sqrt{2}}(-1+i,-1-i,1+i,1-i)^{\top},\label{eq:psi3s-1}\\
\vert\psi_{4}\rangle & =\frac{1}{2\sqrt{2}}(1+i,1-i,-1+i,-1-i)^{\top}.\label{eq:psi4s-1}
\end{alignat}
The states $\vert\psi_{j}\rangle$ ($j=1,..,4$) specify the full
basis for the two-qubit state in $\mathcal{H}_{{\cal P}}\varotimes{\cal H}_{\Pi}$,
where ${\cal H}_{{\cal P}}$ (${\cal H}_{\Pi}$) denote the path (polarization)
Hilbert space.

\begin{table*}[t]
\begin{tabular}{|c||c|c|c|c|}
\hline 
$\mathcal{O}$  & $S_{\text{RR}}$  & $S_{\text{LL}}$  & $S_{\text{RL}}$  & $S_{\text{LR}}$\tabularnewline
\hline 
\hline 
$\sigma_{x}^{{\cal P}}\sigma_{x}^{\Pi}$  & $\frac{1}{\sqrt{2}}\mathbb{I}^{\Pi}$  & $\frac{1}{\sqrt{2}}\sigma_{x}^{\Pi}$  & $-\frac{i}{\sqrt{2}}\sigma_{x}^{\Pi}$  & $\frac{1}{\sqrt{2}}\mathbb{I}^{\Pi}$\tabularnewline
\hline 
$\sigma_{y}^{{\cal P}}\sigma_{z}^{\Pi}$  & $\frac{1}{\sqrt{2}}\sigma_{x}^{\Pi}$  & $\frac{1+i}{2\sqrt{2}}(\sigma_{x}^{\Pi}+\sigma_{y}^{\Pi})$  & $-\frac{1}{\sqrt{2}}\sigma_{y}^{\Pi}$  & $\frac{1+i}{2\sqrt{2}}(\sigma_{x}^{\Pi}-\sigma_{y}^{\Pi})$\tabularnewline
\hline 
$\sigma_{z}^{{\cal P}}\sigma_{y}^{\Pi}$  & $\frac{1}{2\sqrt{2}}(-i\mathbb{I}^{\Pi}+\sigma_{x}^{\Pi}-i\sigma_{y}^{\Pi}+i\sigma_{z}^{\Pi})$  & $\frac{1}{2\sqrt{2}}(\mathbb{I}^{\Pi}-i\sigma_{x}^{\Pi}+\sigma_{y}^{\Pi}+\sigma_{z}^{\Pi})$  & $\frac{1}{2\sqrt{2}}(-\mathbb{I}^{\Pi}-i\sigma_{x}^{\Pi}+\sigma_{y}^{\Pi}-\sigma_{z}^{\Pi})$  & $\frac{1}{2\sqrt{2}}(-i\mathbb{I}^{\Pi}-\sigma_{x}^{\Pi}+i\sigma_{y}^{\Pi}+i\sigma_{z}^{\Pi})$\tabularnewline
\hline 
\end{tabular}

\caption{Components of the unitary 2-qubit gate defined in Eq.~\eqref{eq:Smatrix}
which maps the four eigenstates of the 2-qubit observable $\mathcal{\mathbb{\mathcal{O}}}$
to the four output ports $D_{1}$,..$D_{4}$ in Fig.~\ref{fig2}(d).\label{table3}}

\begin{tabular}{|c||c|c|c|c|c|c|c|c|c|c|c|c|c|c|c|c|c|c|c|c|}
\hline 
$\mathcal{O}$  & $V_{1}$  & $\alpha$  & $\beta$  & $\gamma$  & $\delta$  & $V_{L}$  & $\alpha$  & $\beta$  & $\gamma$  & $\delta$  & $V_{R}$  & $\alpha$  & $\beta$  & $\gamma$  & $\delta$  & $V_{2}$  & $\alpha$  & $\beta$  & $\gamma$  & $\delta$\tabularnewline
\hline 
\hline 
$\sigma_{x}^{{\cal P}}\sigma_{x}^{\Pi}$  & $\sigma_{x}^{\Pi}$  & $0$  & $\frac{\pi}{4}$  & $0$  & $\frac{\pi}{2}$  & $e^{i\frac{\pi}{4}}\sigma_{x}^{\Pi}$  & $0$  & $\frac{\pi}{4}$  & $0$  & $\frac{3\pi}{4}$  & $e^{-i\frac{\pi}{4}}\sigma_{x}^{\Pi}$  & $0$  & $\frac{\pi}{4}$  & $0$  & $\frac{\pi}{4}$  & $\mathbb{I}^{\Pi}$  & $0$  & $0$  & $0$  & $\pi$\tabularnewline
\hline 
$\sigma_{y}^{{\cal P}}\sigma_{z}^{\Pi}$  & $-i\sigma_{z}^{\Pi}$  & $0$  & $0$  & $\frac{\pi}{2}$  & $0$  & $\frac{i}{\sqrt{2}}(\mathbb{\sigma}_{x}^{\Pi}+\sigma_{y}^{\Pi})$  & $-\frac{\pi}{4}$  & $\frac{3\pi}{8}$  & $\frac{\pi}{4}$  & $-\pi$  & $\frac{1}{\sqrt{2}}(\mathbb{\sigma}_{x}^{\Pi}+\sigma_{y}^{\Pi})$  & $-\frac{\pi}{4}$  & $-\frac{\pi}{8}$  & $\frac{\pi}{4}$  & $-\frac{\pi}{2}$  & $\frac{1-i}{2}\mathbb{I}^{\Pi}-\frac{1+i}{2}\sigma_{z}^{\Pi}$  & $-\frac{\pi}{4}$  & $\frac{\pi}{8}$  & $-\frac{\pi}{4}$  & $-\frac{\pi}{4}$\tabularnewline
\hline 
$\sigma_{z}^{{\cal P}}\sigma_{y}^{\Pi}$  & $\tilde{\sigma}_{1}$  & $-\frac{\pi}{4}$  & $-\frac{7\pi}{8}$  & $-\frac{\pi}{2}$  & $\frac{7\pi}{4}$  & $\tilde{\sigma}_{L}$  & $-\frac{\pi}{4}$  & $-\frac{\pi}{4}$  & $\pi$  & $-\frac{3\pi}{4}$  & $\tilde{\sigma}_{R}$  & $-\frac{\pi}{4}$  & $-\frac{\pi}{4}$  & $\frac{\pi}{2}$  & $\frac{3\pi}{4}$  & $\frac{1+i}{2}(\mathbb{\sigma}_{x}^{\Pi}-\mathbb{\sigma}_{y}^{\Pi})$  & $-\frac{\pi}{4}$  & $\frac{\pi}{8}$  & $\frac{\pi}{4}$  & $\frac{3\pi}{4}$\tabularnewline
\hline 
\end{tabular}

\caption{Configuration of the angles of the HWP, QWPs, and phase shifter to
measure the entanglement witnesses defined in Eqs.~\eqref{eq:W1s}
and \eqref{eq:W2s}. For the longer expressions we have introduced
the notation $\tilde{\sigma}_{1}=\frac{1-i}{2}\mathbb{I}^{\Pi}+\frac{1+i}{2}\mathbb{\sigma}_{y}^{\Pi}$,
$\tilde{\sigma}_{L}=\frac{1+i}{2\sqrt{2}}\mathbb{I}^{\Pi}+\frac{1-i}{2\sqrt{2}}(\mathbb{\sigma}_{x}^{\Pi}+\mathbb{\sigma}_{y}^{\Pi}+\mathbb{\sigma}_{z}^{\Pi})$,
and $\tilde{\sigma}_{R}=\frac{1-i}{2\sqrt{2}}\mathbb{I}^{\Pi}+\frac{1+i}{2\sqrt{2}}(-\mathbb{\sigma}_{x}^{\Pi}+\mathbb{\sigma}_{y}^{\Pi}+\mathbb{\sigma}_{z}^{\Pi})$.
\label{table4}}
\end{table*}

\subsubsection{Bell state projection scheme}

We can now define the transformation corresponding to the scheme shown
in Fig.~\ref{fig2}(b):

\begin{equation}
\mathcal{U_{B}}=U_{\text{PBS}}\left[\begin{array}{cc}
U_{\text{\text{QWP}}}(\pi/4) & 0\\
0 & U_{\text{\text{QWP}}}(\pi/4)
\end{array}\right]
\end{equation}
where $U_{\text{PBS}}$ and $U_{\text{\text{QWP}}}(\pi/4)$ are defined
in Eqs.~\eqref{eq:PBS} and \eqref{eq:uqwp45}, respectively. We
find

\begin{alignat}{1}
U_{B}\vert\psi_{1}\rangle & =\frac{e^{-\frac{i\pi}{4}}}{\sqrt{2}}(1,1,0,0)^{\top},\\
\mathcal{U_{B}}\vert\psi_{2}\rangle & =-\frac{e^{-\frac{i\pi}{4}}}{\sqrt{2}}(0,0,1,1)^{\top},\\
\mathcal{U_{B}}\vert\psi_{3}\rangle & =-\frac{e^{-\frac{i\pi}{4}}}{\sqrt{2}}(1,-1,0,0)^{\top},\\
\mathcal{U_{B}}\vert\psi_{4}\rangle & =-\frac{e^{-\frac{i\pi}{4}}}{\sqrt{2}}(0,0,1,-1)^{\top},
\end{alignat}
where the global phase factors are not important here. The detector
$B_{j}$ ($j=1,..4$) in Fig.~\ref{fig2}(b) will thus give a maximum
signal when we are in the Bell state $\vert\psi_{j}\rangle$ ($j=1,..4$)
, while the other three detectors will show a null signal. In our
specific experimental configuration we see that the detectors $B_{1}$
and $B_{2}$ can thus identify the two Bell states $\vert\psi_{1}\rangle$
and $\vert\psi_{2}\rangle$, respectively, as shown in Fig.~\ref{fig2}(c).

\subsubsection{Construction of optimal entanglement witness}

Fidelity-based entanglement witnesses such as those constructed in
Sec.~\ref{sec:ConstructionWitness} can thus be constructed, 
\begin{align}
\mathcal{W}_{1}= & \frac{1}{4}\left[I-\sigma_{x}^{{\cal P}}\sigma_{x}^{\Pi}-\sigma_{y}^{{\cal P}}\sigma_{z}^{\Pi}-\sigma_{z}^{{\cal P}}\sigma_{y}^{\Pi}\right],\label{eq:W1s}\\
\mathcal{W}_{2}= & \frac{1}{4}\left[I-\sigma_{x}^{{\cal P}}\sigma_{x}^{\Pi}+\sigma_{y}^{{\cal P}}\sigma_{z}^{\Pi}+\sigma_{z}^{{\cal P}}\sigma_{y}^{\Pi}\right],\label{eq:W2s}
\end{align}
for $\vert\psi_{1}\rangle,\vert\psi_{2}\rangle$, respectively. These
observables successfully detect entanglement in a white-noise affected
state 
provided the probability of preparing the desired states satisfies
$p>1/3$. 

\subsubsection{Measuring entanglement witnesses using the universal unitary gate}

To implement the entanglement witnesses in Eqs.~\eqref{eq:W1s} and
\eqref{eq:W2s} we employ the universal unitary gate for single-photon
two-qubit states~\citep{englert2001universal} shown in Fig.~\ref{fig2}(d).
The configuration of the universal unitary gate is detailed in Tables
\ref{table3} and \ref{table4}. 

{}
\begin{figure*}
{\includegraphics[width=0.75\textwidth]{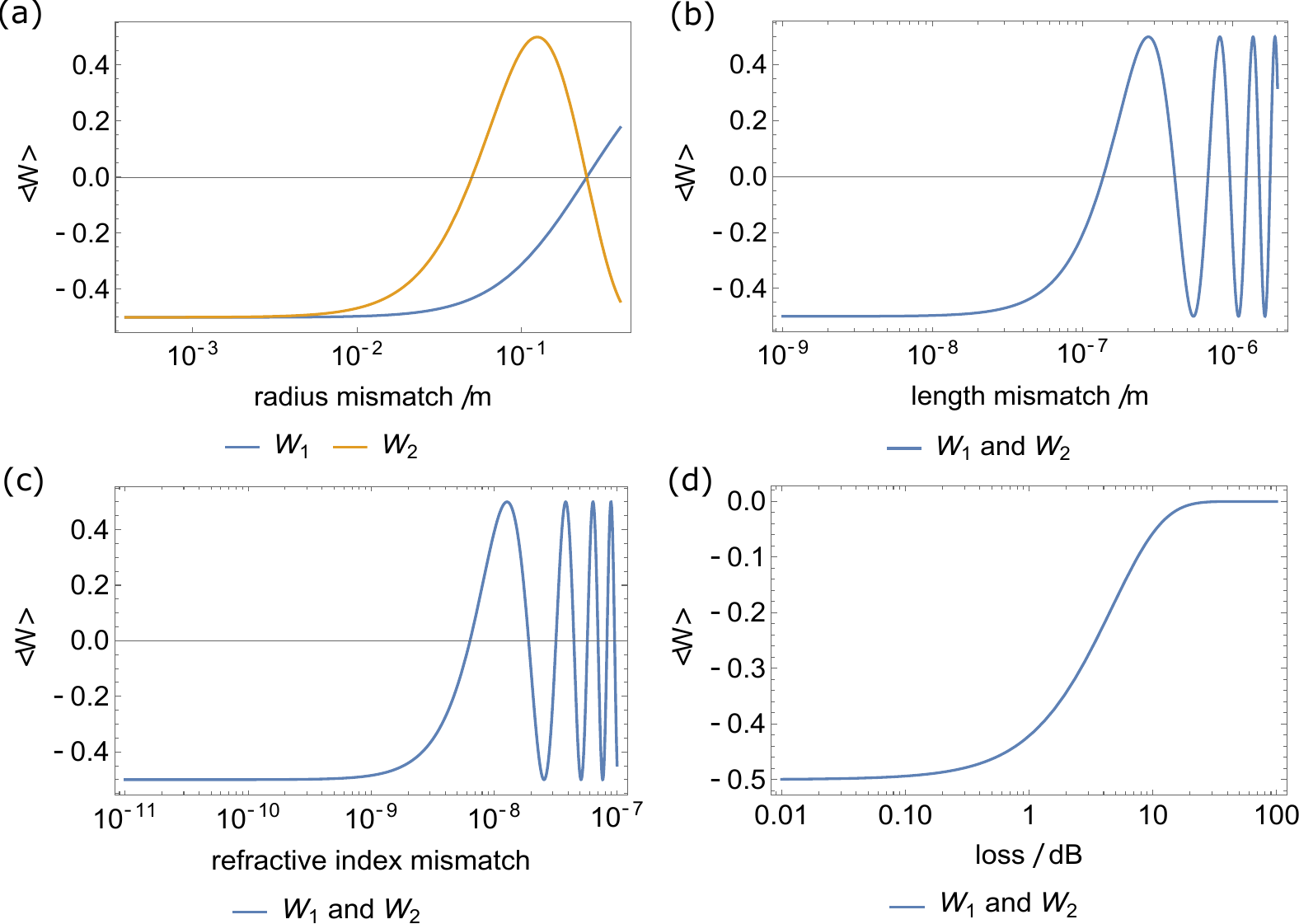}
\caption{{Variation of the expectation values of the entanglement
witnesses $\langle\mathcal{W}_{1}\rangle$ and $\langle\mathcal{W}_{2}\rangle$
due to imperfections and losses for the double-loop scheme. For the
single-loop scheme we find similar estimates for (c), (d), while (a),
(b) become redundant. The unperturbed parameters are: big radius $r_{b}=0.5\text{m},$
small radius $r_{a}=r_{b}/2,$ winding number for big (small) loop
$N_{b}=10$ ($N_{a}=20$) corresponding to the length of the individual
loops of about $\sim31.5\text{m}$, refractive index $n=1.456$, and
photon wavelength $\lambda=800\text{nm}$. (a) Tolerance on the radius
($\delta r$) with pefect length matching. We consider variations
of the radius $r_{b}+\delta r$, but we obtain similar tolerances
for the variation of the radius $r_{a}$. (b) Length mismatch between
the two fiber loops ($\delta L=L_{b}-L_{a}$). (c) Mismatch of the
refractive index for the two polarization ($\delta n=n_{2}-n_{1}$).
(d) Losses in the four fiber paths. Comparable experimental tolerances
have been recently achieved in \cite{experiment}. Given an initial
non-zero mismatch in the refractive index (a), the radius (b), and
the path length (d) one would not require recalibration, but can still
perfrom the experiment as discussed in the main text. This would in
general result in the shifted origin of Figs. 1(c) and 1(e), as the
effect of entanglement generation would still exhibit the same periodiodicity
$4\Omega_{\text{Bell}}$ (see Eqs.~(10) and (11)). }}
\label{fig:robustness}}
\end{figure*}

\subsection{{Robustness estimates}}

{We quantify the effect of imperfections and losses
on the possibility of witnessesing entanglement. We will first focus
on the double-loop scheme presented in Fig.~1, where the optimal entanglement
witnesses are defined in Eqs.~\eqref{eq:W1} and \eqref{eq:W2}.
To estimate the robustness of the double-loop scheme with respect
to the main experimental parameters we introduce the test vector}

{
\begin{alignat}{1}
\vert\tilde{\psi}\rangle=\frac{1}{2}( & e^{-\frac{iL_{a}n_{1}\omega}{c}-\frac{iL_{a}r_{a}\Omega\omega}{c^{2}}},e^{\frac{iL_{a}r_{a}\Omega\omega}{c^{2}}-\frac{iL_{a}n_{2}\omega}{c}},\nonumber \\
 & e^{-\frac{iL_{b}n_{1}\omega}{c}-\frac{iL_{b}r_{b}\Omega\omega}{c^{2}}},e^{\frac{iL_{b}r_{b}\Omega\omega}{c^{2}}-\frac{iL_{b}n_{2}\omega}{c}})^{\top},\label{eq:testv1}
\end{alignat}
for evaluating the expectation value of $\langle\tilde{\psi}\vert\mathcal{W}_{1}\vert\tilde{\psi}\rangle$
and $\langle\tilde{\psi}\vert\mathcal{W}_{2}\vert\tilde{\psi}\rangle$.
The state $\vert\tilde{\psi}\rangle$ takes into account possible
variations from the ideal experimental parameters: $L_{a}$ ($L_{b}$)
denote the total length of the small (big) loops of radius $r_{a}$
($r_{b}$), and $n_{\ensuremath{1}}$($n_{2}$) denote the refractive
index for the slow (fast) axis. We set the unperturbed parameters
to the following values: radius $r_{b}=0.5~\text{m},$winding number
$N_{b}=10$ corresponding to the length of the individual loops of
about $\sim31.5~\text{m}$, refractive index $n=1.456$, and photon
wavelength $\lambda=800~\text{nm}$. We show how $\langle\tilde{\psi}\vert\mathcal{W}_{1}\vert\tilde{\psi}\rangle$
and $\langle\tilde{\psi}\vert\mathcal{W}_{2}\vert\tilde{\psi}\rangle$
depend on imperfections in Fig.~\ref{fig:robustness} (a), (b), and
(c).}

{We now quantify the effect of losses on the optimal
entanglement witness $\mathcal{W}_{1}$ but an analogous analysis
can be performed also for $\mathcal{W}_{2}$ . We set the frequency
to the optimal value $\Omega=\Omega_{\text{Bell}}$ such that the
test vector in Eq.~\eqref{eq:testv1} reduces to}

{
\begin{equation}
\vert\tilde{\psi}\rangle=\frac{1}{2}(-i,i,-1,-1)^{\top}.\label{eq:psit}
\end{equation}
We suppose the loss can occur in any of the four paths with probability
$p$. For example, a single channel loss in the path $a$ and polirazation
$H$ would produce the state}

{
\begin{equation}
\vert\tilde{\psi}_{aH}\rangle=\frac{1}{\sqrt{3}}(0,i,-1,-1)^{\top}.
\end{equation}
with similar definitions for the other three channels $aV$, $bH$,
$bV$. The loss in two channels $(aH,bH)$ would occur with probability
$p^{2}$, resulting in the state}

{
\begin{alignat}{1}
\vert\tilde{\psi}_{aH,aV}\rangle & =\frac{1}{\sqrt{2}}(0,0,-1,-1)^{\top},
\end{alignat}
with analogous definitions for the chanels $(aH,bH)$, $(aH,bV)$,
$(aV,bH)$, $(aV,bV)$, $(bH,bV)$. Finally, the three channel loss
$(aH,aV,bH)$ would occur with probability $p^{3}$, with the associated
state given by:}

{
\begin{alignat}{1}
\vert\tilde{\psi}_{aH,aV,bH}\rangle & =(0,0,0,-1)^{\top},\label{eq:psit2}
\end{alignat}
with again similar definitions for the other cases $(aH,aV,bV)$,
$(aH,bH,bV)$, $(aV,bH,bV)$. To compute the expectation value of
the entanglement witness we need to average over the four types of
states defined in Eqs.~\eqref{eq:psit}-\eqref{eq:psit2}; no loss
with probability $1-(4p+6p^{2}+4p^{3})/14$, single chanel loss with
probability $p/14$ ($4$ different states), two chanel loss with
probability $p^{2}/14$ ($6$ different states), and three chanel
loss with probability $p^{3}/14$ ($4$ different states). In other
words, we construct the density matrix:}

{
\begin{alignat}{1}
\rho= & (1-\frac{4p+6p^{2}+4p^{3}}{14})\vert\tilde{\psi}\rangle\langle\tilde{\psi}\vert+\frac{p}{14}\sum_{j}\vert\tilde{\psi}_{j}\rangle\langle\tilde{\psi}_{j}\vert\nonumber \\
 & +\frac{p^{2}}{14}\sum_{j,l}\vert\tilde{\psi}_{jl}\rangle\langle\tilde{\psi}_{jl}\vert+\frac{p^{3}}{14}\sum_{j}\vert\tilde{\psi}_{jlm}\rangle\langle\tilde{\psi}_{jlm}\vert
\end{alignat}
where $j,l,m$ denote a loss chanel $aH,aV,bH,bV$. We show how the
value of $\langle\mathcal{W}_{j}\rangle=\text{tr}[\rho\,\mathcal{W}_{j}]$
($j=1,2$) is attenuated in Fig. 1(d).}

{An similar analysis of imperfections and losses, discussed
above for the double-loop scheme, can be also applied to the single-loop
scheme shown in Fig.~\ref{fig2}. We consider the constructed entanglement
witnesses in Eqs.~(\ref{eq:W1s}) and (\ref{eq:W2s}) and introduce
the test vector 
\begin{alignat}{1}
\vert\tilde{\psi}\rangle=\frac{1}{2}( & e^{-\frac{iLr\omega\Omega}{c^{2}}-\frac{iLn_{1}\omega}{c}},e^{\frac{iLr\omega\Omega}{c^{2}}-\frac{iLn_{2}\omega}{c}}\nonumber \\
 & e^{\frac{iLr\omega\Omega}{c^{2}}-\frac{iLn_{1}\omega}{c}},e^{-\frac{iLr\omega\Omega}{c^{2}}-\frac{iLn_{2}\omega}{c}})^{\top},\label{eq:testvS}
\end{alignat}
which is used for evaluating the expectation values, $\langle\tilde{\psi}\vert\mathcal{W}_{1}\vert\tilde{\psi}\rangle$
and $\langle\tilde{\psi}\vert\mathcal{W}_{2}\vert\tilde{\psi}\rangle$.
We find that the tolerance on the mismatch of the refractive index
for the two polarizations ($\delta n=n_{2}-n_{1}$) is comparable
to the computed for the double loop scheme in Fig.~\ref{fig:robustness}(c),
and a similar attenuation of the entanglement witness with losses
to the one shown Fig.~\ref{fig:robustness}(d). On the other hand,
the single-loop scheme is no longer susceptible to radius and length
mismatch, as the scheme relies on a single fiber loop, and hence Fig.~\ref{fig:robustness}(a)
and Fig.~\ref{fig:robustness}(b) become redundant. }
\end{document}